\documentclass{JINST}
\usepackage{lineno}

\title{Electromagnetic response of a highly granular hadronic calorimeter}

\author{\centering 
\LARGE\bf The CALICE Collaboration
}

\author{\centering
C.\,Adloff, 
J.\,Blaha, 
J.-J.\,Blaising, 
C.\,Drancourt,
A.\,Espargili\`{e}re, 
R.\,Gaglione, 
N.\,Geffroy, 
Y.\,Karyotakis, 
J.\,Prast,
G.\,Vouters
\\ \it
Laboratoire d'Annecy-le-Vieux de Physique des Particules, Universit\'{e} de Savoie,
CNRS/IN2P3,
9 Chemin de Bellevue BP110, F-74941 Annecy-le-Vieux CEDEX, France
}

\author{\centering
K.\,Francis,
J.\,Repond, 
J.\,Smith\footnote{Also at University of Texas, Arlington},
L.\,Xia 
\\ \it
Argonne National Laboratory,
9700 S.\ Cass Avenue,
Argonne, IL 60439-4815,
USA}

\author{\centering
E.\,Baldolemar, 
J.\,Li\footnote{Deceased}, 
S.\,T.\,Park, 
M.\,Sosebee, 
A.\,P.\,White, 
J.\,Yu
\\ \it
Department of Physics, SH108, University of Texas, Arlington, TX 76019, USA
}

\author{\centering
T.\,Buanes, G.\,Eigen
\\ \it
University of Bergen, Inst.\, of Physics, Allegaten 55, N-5007 Bergen, Norway
}

\author{\centering
Y.\,Mikami, 
N.\,K.\,Watson 
\\ \it
University of Birmingham,
School of Physics and Astronomy,
Edgbaston, Birmingham B15 2TT, UK
}

\author{\centering 
T.\,Goto, 
G.\,Mavromanolakis\footnote{Now at CERN}, 
M.\,A.\,Thomson, 
D.\,R.\,Ward
W.\,Yan\footnote{Now at Dept.\, of Modern Physics, Univ. of Science and Technology of China, 96 Jinzhai Road, Hefei, Anhui, 230026, P.\, R.\, China}
\\ \it
University of Cambridge, Cavendish Laboratory, J J Thomson Avenue, CB3 0HE, UK
}

\author{\centering 
D.\,Benchekroun, 
A.\,Hoummada, 
Y.\,Khoulaki
\\ \it
Universit\'{e} Hassan II A\"{\i}n Chock, Facult\'{e} des sciences.\, B.P. 5366 Maarif, Casablanca, Morocco
}

\author{\centering
M.\,Benyamna, 
C.\,C\^{a}rloganu, 
F.\,Fehr, 
P.\,Gay, 
S.\,Manen, 
L.\,Royer
\\ \it
Clermont Univertsit\'e, Universit\'e Blaise Pascal, CNRS/IN2P3, LPC, BP
10448, F-6300ls
0 Clermont-Ferrand, France
}

\author{\centering
G.\,C.\,Blazey,
A.\,Dyshkant, 
J.\,G.\,R.\,Lima, 
V.\,Zutshi
\\ \it
NICADD, Northern  Illinois University,
Department of Physics,
DeKalb, IL 60115,
USA
}

\author{\centering 
J.\,-Y.\,Hostachy, 
L.\,Morin
\\ \it
Laboratoire de Physique Subatomique et de Cosmologie - Universit\'{e} Joseph Fourier Grenoble 1 -
CNRS/IN2P3 - Institut Polytechnique de Grenoble,
53, rue des Martyrs,
38026 Grenoble CEDEX, France
}

\author{\centering 
U.\,Cornett, 
D.\,David, 
R.\,Fabbri, 
G.\,Falley, 
K.\,Gadow, 
E.\,Garutti,
P.\,G\"{o}ttlicher, 
C.\,G\"{u}nter,
S.\,Karstensen, 
F.\,Krivan,
A.\,-I.\,Lucaci-Timoce\footnote{Now at CERN}, 
S.\,Lu, 
B.\,Lutz, 
I.\,Marchesini, 
N.\,Meyer,
S.\,Morozov, 
V.\,Morgunov\footnote{On leave from ITEP}, 
M.\,Reinecke, 
F.\,Sefkow, 
P.\,Smirnov,
M.\,Terwort,
A.\,Vargas-Trevino, 
N.\,Wattimena, 
O.\,Wendt
\\ \it
DESY, Notkestrasse 85,
D-22603 Hamburg, Germany
}

\author{\centering  
N.\,Feege, 
J.\,Haller, 
S.\,Richter, 
J.\,Samson
\\ \it
Univ. Hamburg,
Physics Department,
Institut f\"ur Experimentalphysik,
Luruper Chaussee 149,
22761 Hamburg, Germany
}

\author{\centering 
P.\,Eckert,
A.\,Kaplan,
 H.\,-Ch.\,Schultz-Coulon,
 W.\,Shen,
 R.\,Stamen,
 A.\,Tadday
\\ \it
 University of Heidelberg, Fakultat fur Physik und Astronomie,
Albert Uberle Str. 3-5 2.OG Ost,
D-69120 Heidelberg, Germany
}

\author{\centering 
B.\,Bilki, E.\,Norbeck, 
Y.\,Onel
\\ \it
University of Iowa, Dept. of Physics and Astronomy,
203 Van Allen Hall, Iowa City, IA 52242-1479, USA
}

\author{\centering 
G.\,W.\,Wilson
\\ \it
University of Kansas, Department of Physics and Astronomy,
Malott Hall, 1251 Wescoe Hall Drive, Lawrence, KS 66045-7582, USA
}

\author{\centering 
K.\,Kawagoe,  
S.\,Uozumi\footnote{Now at Kyungpook National University.}
\\ \it
 Department of Physics, Kobe University, Kobe, 657-8501, Japan
}

\author{\centering 
J.\,A.\,Ballin, 
P.\,D.\,Dauncey, 
A.\,-M.\,Magnan,
H.\,S.\,Yilmaz, 
O.\,Zorba
\\ \it
Imperial College London, Blackett Laboratory,
Department of Physics,
Prince Consort Road,
London SW7 2AZ, UK 
}

\author{\centering 
V.\,Bartsch\footnote{Now at University of Sussex, Physics and
  Astronomy Department, Brighton, Sussex, BN1 9QH, UK}, 
M.\,Postranecky, M.\,Warren, M.\,Wing
\\ \it
Department of Physics and Astronomy, University College London,
Gower Street,
London WC1E 6BT, UK
}

\author{\centering 
F.\,Salvatore\footnotemark[8]
\\ \it
Royal Holloway University of London,
Dept. of Physics,
Egham, Surrey TW20 0EX, UK
}

\author{\centering 
E.\,Calvo~Alamillo, 
M.-C.\, Fouz, 
J.\,Puerta-Pelayo 
\\ \it
CIEMAT, Centro de Investigaciones Energeticas, Medioambientales y Tecnologicas, Madrid, Spain 
}

\author{\centering 
V.\,Balagura, 
B.\,Bobchenko, 
M.\,Chadeeva, 
M.\,Danilov, 
A.\,Epifantsev, 
O.\,Markin, 
R.\,Mizuk, 
E.\,Novikov, 
V.\,Rusinov, 
E.\,Tarkovsky
\\ \it
Institute of Theoretical and Experimental Physics, B. Cheremushkinskaya ul. 25,
RU-117218 Moscow, Russia
}

\author{\centering 
V.\,Kozlov, 
Y.\,Soloviev 
\\ \it
P.\,N.\, Lebedev Physical Institute,
Russian Academy of Sciences,
117924 GSP-1 Moscow, B-333, Russia
}

\author{\centering 
P.\,Buzhan, B.\,Dolgoshein, A.\,Ilyin, V.\,Kantserov, V.\,Kaplin, A.\,Karakash, E.\,Popova, S.\,Smirnov 
\\ \it
Moscow Physical Engineering Inst., MEPhI,
Dept. of Physics,
31, Kashirskoye shosse,
115409 Moscow, Russia
}

\author{\centering 
A.\,Frey\footnote{Now at University of G\"{o}ttingen}, 
C.\,Kiesling,
K.\,Seidel, 
F.\,Simon,
C.\,Soldner, 
L.\,Weuste
\\ \it
Max Planck Inst. f\"ur Physik,
F\"ohringer Ring 6,
D-80805 Munich, Germany
}

\author{\centering 
J.\,Bonis, 
B.\,Bouquet,    
S.\,Callier, 
P.\,Cornebise, 
Ph.\,Doublet,
F.\,Dulucq, 
M.\,Faucci Giannelli, 
J.\,Fleury,
G.\,Guilhem, 
H.\,Li,  
G.\,Martin-Chassard, 
F.\,Richard, 
Ch.\,de la Taille, 
R.\,P\"{o}schl, 
L.\,Raux,  
N.\,Seguin-Moreau, 
F.\,Wicek
\\ \it
Laboratoire de L'acc\'elerateur Lin\'eaire,
Centre d'Orsay, Universit\'e de Paris-Sud XI,
BP 34, B\^atiment 200,
F-91898 Orsay CEDEX, France
}

\author{\centering 
M.\,Anduze,
V.\,Boudry, 
J-C.\,Brient, 
D.\,Jeans, 
P.\,Mora de Freitas, 
G.\,Musat, 
M.\,Reinhard, 
M.\,Ruan,  
H.\,Videau
\\ \it
      Laboratoire Leprince-Ringuet (LLR)  -- \'{E}cole Polytechnique,
      CNRS/IN2P3,
      Palaiseau, F-91128 France
}

\author{\centering 
B.\,Bulanek,
J.\,Zacek 
\\ \it
Charles University, Institute of Particle \& Nuclear Physics,
V Holesovickach 2,
CZ-18000 Prague 8, Czech Republic  
}

\author{\centering 
J.\,Cvach, 
P.\,Gallus, 
M.\,Havranek, 
M.\,Janata, 
J.\,Kvasnicka,
D.\,Lednicky,
M.\,Marcisovsky, 
I.\,Polak, 
J.\,Popule, 
L.\,Tomasek, 
M.\,Tomasek, 
P.\,Ruzicka, 
P.\,Sicho, 
J.\,Smolik, 
V.\,Vrba, 
J.\,Zalesak 
\\ \it
Institute of Physics, Academy of Sciences of the Czech Republic, Na Slovance 2,
CZ-18221 Prague 8, Czech Republic
}

\author{\centering 
B.\,Belhorma,
H.\,Ghazlane
\\ \it
Centre National de l'Energie, des Sciences et des Techniques Nucl\'{e}aires, 
B.P. 1382, R.P. 10001, Rabat, Morocco
}

\author{\centering              
K.\,Kotera, M.\, Nishiyama, T.\,Takeshita, S.\,Tozuka
\\ \it
Shinshu Univ.\,,
Dept. of Physics,
3-1-1 Asaki,
Matsumoto-shi, Nagano 390-861,
Japan
}

\abstract{The CALICE collaboration is studying the design of high
  performance electromagnetic and hadronic calorimeters for future
  International Linear Collider detectors. For the hadronic
  calorimeter, one option is a highly granular
  sampling calorimeter with steel as absorber and scintillator layers as active material.
  High granularity is obtained by segmenting the scintillator into small tiles 
  individually read out via silicon photo-multipliers (SiPM).
  A prototype has been built, consisting of thirty-eight sensitive layers, segmented into about eight thousand channels. 
  In 2007 the prototype was exposed to positrons and hadrons using the CERN SPS beam,
  covering a wide 
  range of beam energies and angles of incidence. The challenge of cell equalization and 
  calibration of such a large number of channels is best validated using electromagnetic processes. 
  The response of the prototype steel-scintillator calorimeter, including linearity and uniformity, to electrons is
  investigated and described.
  }


\begin{document}

\section{Introduction}\label{sec:intro}
A new generation of calorimeters that exploit unprecedented high granularity 
to reach excellent jet energy resolution is one of the main R\&D goals towards 
the future International Linear Collider (ILC)~\cite{RDR_machine}.
The particle flow (PFLOW~\cite{PFA2,PFA3,Thomson}) algorithm favors single particle 
separability over single particle energy resolution in the attempt to improve the overall
jet energy resolution. Typical single hadronic showers in the 10--100 GeV range 
are best separated in a hadronic calorimeter with cell size of the order of $3\times 3$ cm$^2$~\cite{Thomson}.
In addition, fine longitudinal segmentation is required for PFLOW algorithms to be effective.

The CALICE collaboration~\cite{CALICE} is studying several
calorimeter designs for experiments at the ILC. 
With the first generation of prototype detectors new readout technologies have been established 
for highly granular calorimeters and the stability of 
these detectors has been demonstrated. 
Furthermore, a unique set of data has been collected to 
study hadronic showers  at low and medium energies in detail with high resolution
longitudinal and transverse sampling.  

This paper focuses on the prototype of an analog hadron calorimeter (AHCAL)
consisting of 38 layers of highly-segmented scintillator plates 
sandwiched between  2\,cm thick steel plates.
Each scintillator tile is an individual calorimeter cell read out by
a silicon photo-multiplier (SiPM). Details on the calorimeter structure, calibration and readout electronics are given in Section~\ref{sec:proto}.

Tests using particle beams have been conducted in order to evaluate the performance of the highly granular calorimeters built by CALICE. 

In 2007, the whole detector with 38 active layers was
commissioned and exposed to muon, positron and pion beams in
the energy range  6\,GeV to 80\,GeV provided by the CERN Super Proton Synchrotron (SPS~\cite{CERNbeamline}), on the H6
beam line. In 2008, the AHCAL together with
the ECAL and TCMC were moved to the FNAL Meson Test Beam Facility (MTBF~\cite{FNALbeamline}) to take data in the 1--6~GeV energy
range over the course of two years.

Ongoing data analyses will quantify the energy and spatial resolutions of the prototype for hadrons, 
and will continue the validation and further development of existing models of hadronic showers, e.\,g. the various {\sc Geant}4~\cite{G4} physics lists. They will also be important for the experimental validation of the PFLOW approach~\cite{Oleg}.  
The studies in this article focus on the calibration and performance 
of the device when exposed to electrons and positrons.

In Section \ref{sec:proto}, the AHCAL is described, and in Section \ref{sec:calibration} the electromagnetic calibration procedure is discussed. The CERN test beam experiment is described in Section \ref{sec:testbeam}. Results on calorimeter response to positrons are given in Section \ref{sec:EMresponse}, followed by uniformity studies in Section \ref{sec:uniformity}. Conclusions are reported in Section \ref{sec:conclusion}.


\section{Prototype calorimeter}
\label{sec:proto}

\begin{figure}[!t]
  \begin{center}
    \includegraphics[width=0.45\textwidth]{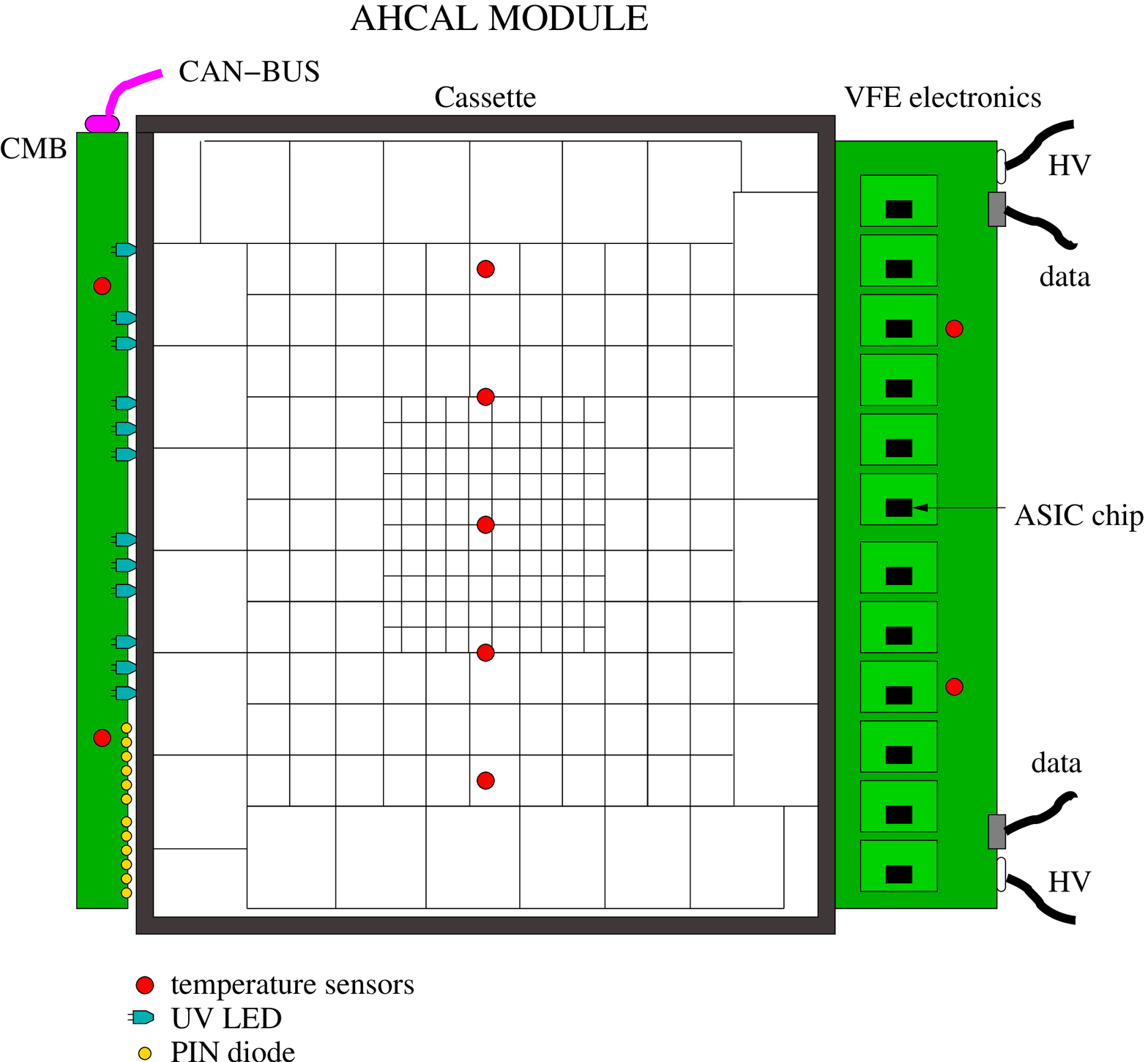}
    \includegraphics[width=0.45\textwidth]{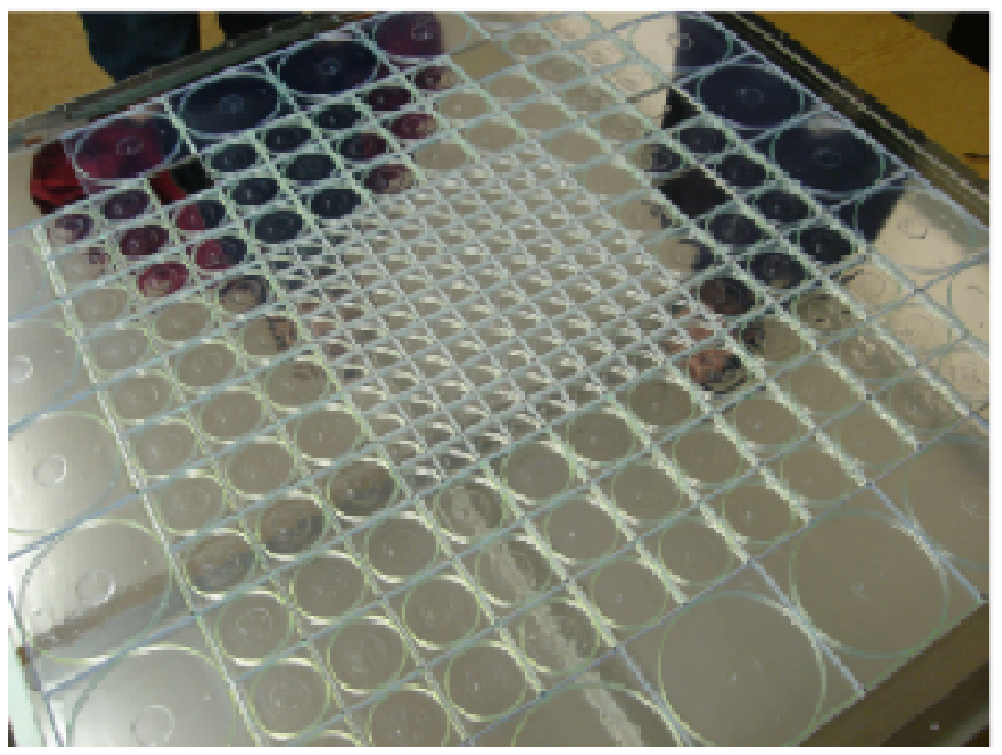}
 \end{center}
 \caption[]{{Sketch of one AHCAL module (left). The scintillator tiles with SiPM
     readout are embedded in a steel cassette. The SiPM
     signal is routed to the VFE electronics located on the right
     side. The calibration and monitoring board (CMB), located on the left
     side, provides UV LED light for calibration and the CAN-BUS
     readout for temperature sensors located inside the cassette and
     on the electronics (red dots). Picture of one active layer of the CALICE AHCAL
     prototype (right). }}
  \label{fig:module}
\end{figure}

The AHCAL is a sampling calorimeter with alternating 2\,cm thick steel plates and highly-segmented scintillator-based active layers. The single calorimeter cell is a scintillator tile read out via a SiPM.
The scintillator tiles are 0.5\,cm thick and have a size of 3\,cm$\times$3\,cm in the 30\,cm$\times$30\,cm  core region and increase to 6\,cm$\times$6\,cm and 12\,cm$\times$12\,cm in the rings surrounding the core. A sketch of one AHCAL module, as well as a picture of an open module showing the arrangement of scintillator tiles, are shown in Figure~\ref{fig:module}.
A wavelength shifting fiber is embedded in the tile, which collects the scintillation light and guides it to the SiPM. The other fiber end is pressed against a 3M reflector foil. The details of one core tile are shown more clearly later in this paper, in the left side of Figure~\ref{fig:homogeneity}. 
The four sides of each tile are matted by a chemical treatment providing a white surface that serves as a diffuse reflector. The two large faces of the tile are not individually coated, instead a large 3M reflector foil is glued to each side of the metal cassette hosting all tiles providing reflectivity via an air contact. The imperfect reflective coating of the tile edges is responsible for about 2.5\,\% light cross-talk between neighboring cells of 3~cm edge size. 

SiPM devices from the MEPhI/PULSAR group have been used, which have
an active area of 1.1\,mm$\times$\,1.1 mm containing 1156 pixels, each 32\,$\mu$m$\times$32\,$\mu$m in size. 
SiPMs are operated with a reverse bias voltage of $\sim$50 V, which lies a few volts above the breakdown voltage, resulting in a gain of $\sim$10$^6$. A poly-silicon quenching resistor on each pixel is used to quench the Geiger discharge. The resistor values vary between 0.5\,M$\Omega$ and 20\,M$\Omega$ for the various batches of SiPM produced. Larger resistor values have been favored as they yield a longer pixel recovery time up to 1\,$\mu$s. In this way a pixel cannot be fired multiple times during one scintillator light pulse, though making it easier to monitor the SiPM response curve with LED light. More details on the SiPM working principle and its properties are given in~\cite{sipm,sipm3}.

The active layers are referred to as modules, and the sum of active and passive material adds up to a total depth of 5.3 nuclear interaction lengths ($\lambda_i$). 
A more detailed description of the AHCAL prototype structure is given in~\cite{ahcal}.
The analog SiPM signal is routed to the very-front-end (VFE)
electronics where a dedicated ASIC chip~\cite{ASIC} is used for multiplexed readout of 18 SiPMs.
The integrated components of the ASIC chip allow to select one of sixteen
fixed preamplifier gain factors from 1 to 100 mV/pC, and one of sixteen CR--RC$^2$ shapers with
peaking times from 40 to 180 ns.

Since the AHCAL was the first detector to employ such a large number of SiPMs, a specialized system for monitoring the long-term stability and performance of the photodetectors was required. 
In order to monitor the SiPM response function in-situ, a versatile UV LED light distribution system was developed~\cite{nanda_diploma}. A calibration and monitoring board (CMB) connected to each module distributes UV light from an LED to each tile via clear fibers. The LEDs are pulsed with 10 ns wide signals steerable in amplitude. By varying the voltage, the LED intensity covers the full dynamic range from zero to saturation (about 70 times the signal of a minimum-ionizing particles).
Furthermore, the LED system monitors variations of SiPM gain and signal response, both sensitive to temperature and voltage fluctuations. 
The LED light itself is monitored with a PIN photo-diode to correct for fluctuations in the LED light intensity.

\section{Calibration procedure}
\label{sec:calibration}

One of the aims of the tests discussed here is to establish a reliable and
robust calibration chain. This requires measurements with beam particles and with light 
from the LED monitoring system. 
The calibration chain is summarized in the following steps:
\begin{itemize}
\item calibration of the cell response and cell-to-cell equalization;
\item monitoring of the SiPM gain and corrections for the non-linear
 response;
\item calibration to an energy scale (in GeV) with electromagnetic showers.
\end{itemize}

The cell-by-cell calibration, and with that the equalization of all cell responses, is performed using minimum-ionizing particles (MIPs) provided by a broad muon beam with an approximately Gaussian profile with a width of about 30--40 cm, illuminating all cells in the detector. For each cell, a calibration factor, $C_i^{\rm{MIP}}$, is determined from the most probable value of the measured energy spectrum for muons in ADC units, which is extracted with a fit using a Landau function convolved with a Gaussian. This fit accounts for the distribution of energy loss of muons in the scintillator tiles as well as for contributions from photon counting statistics and electronic noise. The combined systematic and statistical uncertainty for these fits was typically on the order of 2\%. The muons are generally parallel to the beamline and perpendicular to the detector front face. In this way all cells can be calibrated at the same time, minimizing the impact of temperature induced variations. 

The SiPM gain and photo-detection efficiency are temperature dependent. The product of the two determines the SiPM response, which typically decreases by 3.7\%/K. A procedure has been developed to correct temperature-induced variations in the calorimeter response using temperature measurements in each module. This procedure and its stability will be described in more detail in~\cite{muon_paper}. For the analysis presented in this paper, data samples have been selected to cover a temperature range of less than 0.5~K to reduce the impact of such corrections.  To  account for the included temperature variations, the visible energy of each data set is scaled by -3.7\%/K to the average temperature of the muon data used for calibration.

The number of SiPM pixels, $A_i[{\rm pix}]$, firing for a single cell $i$
is related to the ADC value for the cell, $A_i[{\rm ADC}]$, and the corresponding SiPM gain, $C_i^{\rm pix} [{\rm ADC}]$ by
$A_i\,[{\rm pix}] = A_i\,[{\rm ADC}]/C_i^{\rm pix}[{\rm ADC}]$.
The procedure
to obtain the gain of each individual SiPM is discussed in
Section~\ref{sec:gain_calib}.

The limited number of SiPM pixels leads to a non-linear response for large signals.
These effects are corrected for by a function, $f_{\rm{sat}}(A [{\rm pix}])$, depending on the number of fired pixels $A [{\rm pix}]$. 
This procedure is discussed in detail in Section~\ref{sec:saturation}.

Finally, a common calibration factor, $w$, scales the visible energy of electrons in each
cell in units of MIP to the total deposited energy in units of GeV.
This factor is determined to be $w = (42.3 \pm 0.4)$\,MIP/GeV, as
discussed in Section~\ref{sec:EMresponse}. 

Therefore, in summary, the reconstructed energy of electromagnetic showers in the calorimeter is expressed as
\begin{equation}
  \label{eq:esum}
	E_{\rm{reco}} [{\rm GeV}] = \frac{\sum_iE_i [{\rm MIP}]}{w [{\rm MIP/GeV}]},
\end{equation}
where the energy of one single cell with index $i$ is $E_i$. 
The energy, $E_i$, given in units of MIP is calculated according to
\begin{equation}
  \label{eq:calib}
  E_i\,[{\rm MIP}] =  \frac{A_i\,[{\rm ADC}]}{C_i^{\rm{MIP}}} \cdot
  f_{\rm{sat}}(A_i\,[{\rm pix}]).
\end{equation}

\subsection{SiPM gain and electronics inter-calibration factors}\label{sec:gain_calib}
The gain of each individual SiPM is extracted from single photoelectron
spectra taken in dedicated runs with low LED light intensity. 
LED light is necessary as the best determination of the gain requires 
a single photoelectron spectrum with a Poisson mean of about 1.5 p.e. 
and the mean obtained from dark noise events is below 0.5 p.e.

\begin{figure}[!t]
  \begin{center}
    \includegraphics[width=0.6\textwidth]{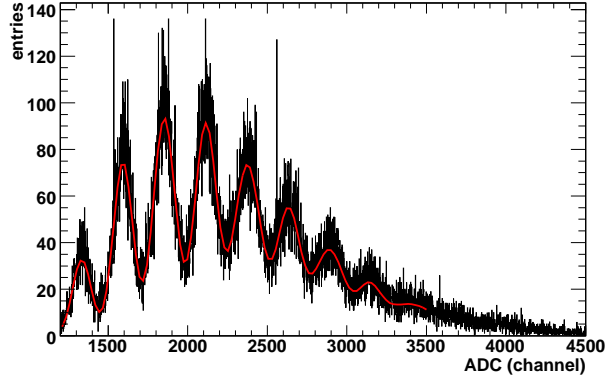}
  \end{center}
  \caption[]{{Single photoelectron peak spectrum taken with a SiPM in
      the AHCAL detector.}}
  \label{gain_spec}
\end{figure}

The SiPM gain, $G_i^{\rm{fit}}$,
is the distance between two
consecutive peaks in the single photoelectron spectrum. 
A typical gain spectrum is shown in Figure~\ref{gain_spec}. A
multi-Gaussian fit is performed to the single photoelectron peaks to
determine their average relative distance~\cite{beni_diploma}. 
The mean of each Gaussian function in the multi-Gaussian sum 
is left as a free parameter. Before fitting, a peak finder routine is used 
to set each peak mean value to the approximate location of the corresponding 
photoelectron peak. It has been seen that fixing the distance between peaks to one common parameter 
reduced significantly the number of converged fits.  
The width of each Gaussian function is dominated by electronic
noise, but for large number of pixels fired 
the statistical contribution becomes visible, which lead to an increase of the peak
width. Accordingly, the width of each peak is left as a free parameter.
Finally, the SiPM gain is defined from the fit result as the distance between pedestal (peak zero) and second peak divided by two. An additional consistency check is performed to ensure the distance between pedestal and first peak agrees with the defined gain value at better than 2\%. Gaussian fits to peaks higher than the 3rd one are not directly used in determining the gain, but their proper description helps to improve the stablity of the fits and to avoid bias on the peak position.
The uncertainty on the gain determination is mainly due to the fit and
is about 2\,\% for fits which pass quality criteria.


SiPM gain measurements were repeated approximately every eight hours during 
test beam operation. The SiPM gain varies with temperature 
and the gain measurements can be used to stabilize the calorimeter response over time.
The temperature dependence of the SiPM gain is further discussed in~\cite{muon_paper}.
The efficiency of the gain extraction is defined as the number of successful 
fits in one gain run divided by the number of channels which can be calibrated. 
About 2\,\% of all SiPMs are considered inactive because of initially bad soldering
or subsequent broken connections to the SiPM leads. 
Additionally, about 0.11\,\% of all channels are connected to a broken LED. All these 
channels are not accounted for in the total of channels that can be calibrated. 

\begin{figure}[!t]
  \begin{center}
    \includegraphics[width=0.49\textwidth]{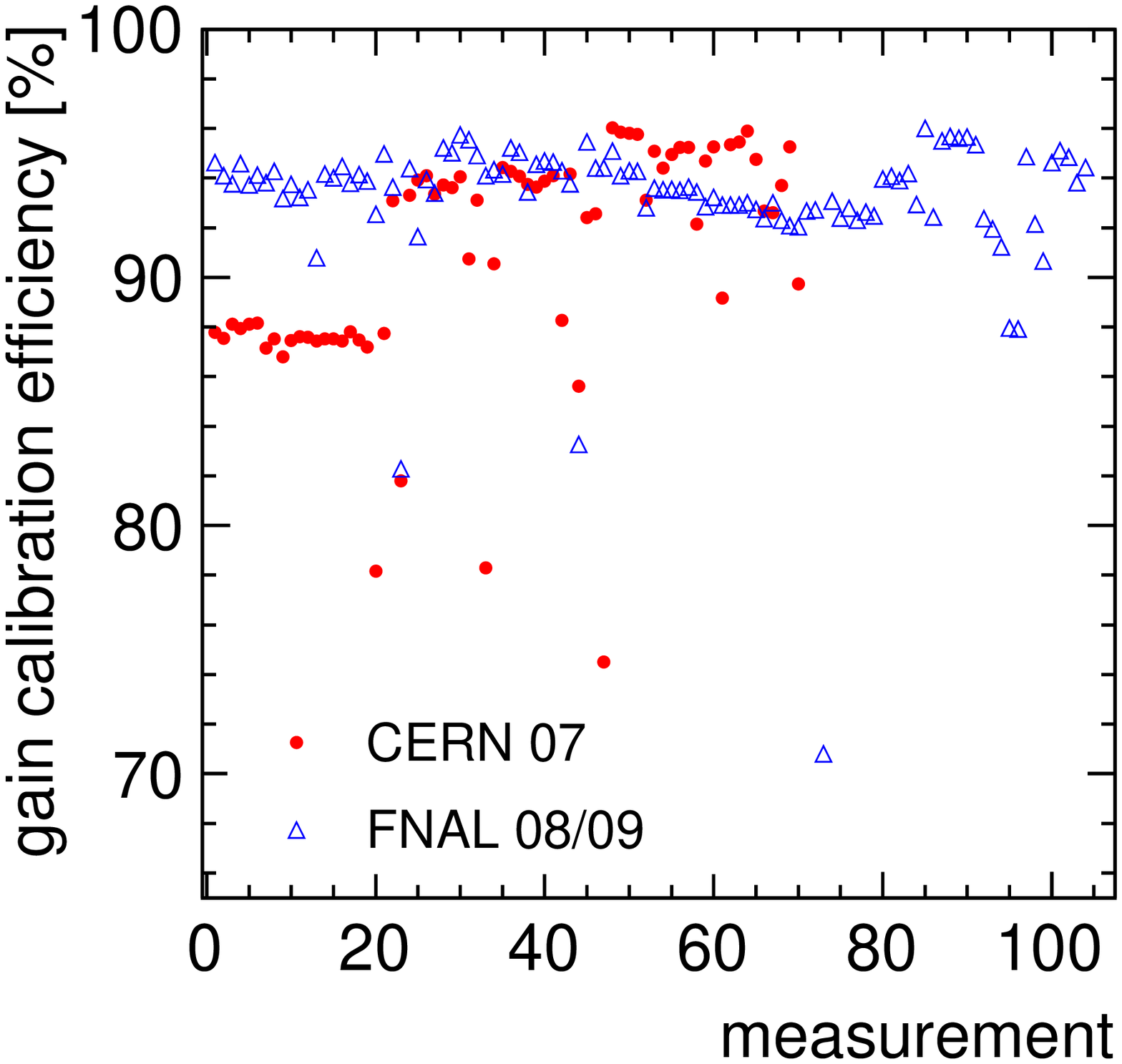}
    \includegraphics[width=0.49\textwidth]{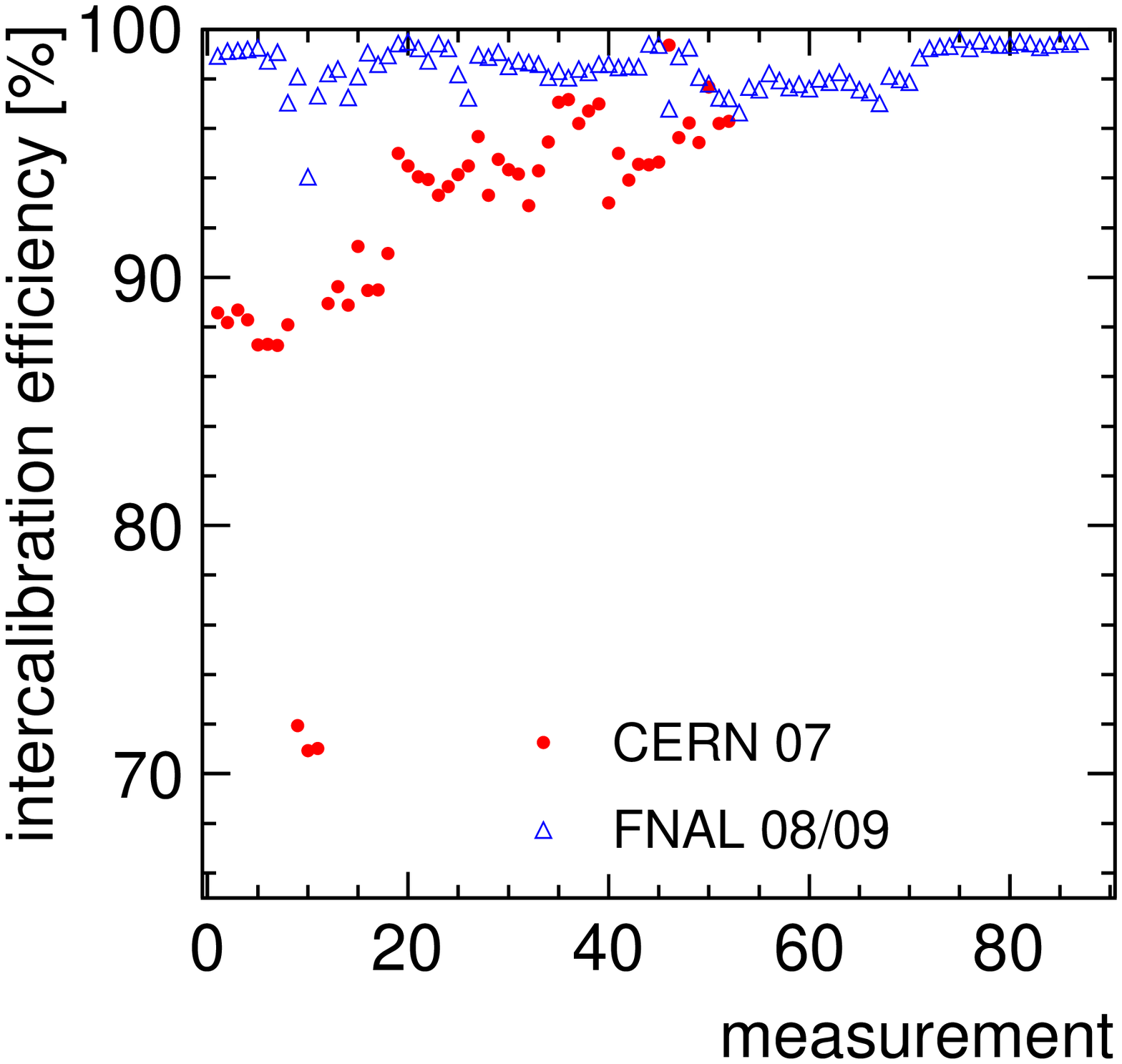}
  \end{center}
  \caption[]{Gain calibration efficiency (left) and  electronics
      inter-calibration efficiency (right) over the AHCAL data taking period
      at CERN in 2007 (red dots) and at FNAL in 2008 (open blue
      triangles). More than 85.0\,\% of the channels could be
      monitored for gain and inter-calibration variation during these
      periods.  }
  \label{gain_eff}
\end{figure}

The efficiency of the gain extraction with one measurement run is indicative 
of the quality of the LED monitoring system, namely the small spread of LED light intensity.
Figure~\ref{gain_eff} shows the efficiency of the gain extraction for a
series of runs taken in the first three months of data taking at CERN and in the first three months
at FNAL. Initial problems during the system commissioning phase led to low efficiency,
but after commissioning a gain extraction efficiency of about 95\,\% per run has been achieved.  The gain
efficiency was also stable 
after transportation and throughout the FNAL runs.
Combination of several gain runs yields calibration of more than 99\,\% of all cells. The remaining 1\,\% of cells are calibrated with the average of the module to which they belong.

The measurement of SiPM gain is performed with a special mode of the readout chip, with a 
choice of high pre-amplification gain and short peaking time of 50~ns
which improves the signal to noise ratio at the single pixel level. 
In contrast, the muon calibration and the physics data taking are performed with
approximately ten times smaller electronic amplification, to optimally fit
the available dynamic range, and about 180~ns peaking time to 
provide sufficient latency for the beam trigger. 
The inter-calibration factor, $I_i$, of the chip gain between the calibration mode (CM) and the physics data mode (PM) along
with the SiPM gain are used to determine the overall SiPM calibration factor, $C_i^{\rm{pix}} [{\rm ADC}]$, used in Eq.~\ref{eq:calib}:
\begin{equation}
  \label{eq:gain}
  C_i^{\rm{pix}} =  G_i^{\rm{fit}}[{\rm ADC(CM)}] / I_i.
\end{equation}

The extraction of the inter-calibration coefficients depends on 
the linear response of the chip in both modes for an overlapping range of input signals. 
The input signal is
provided by the LED system injecting light into the tiles. The amplitude
of the signal is varied within the linear range by varying the LED
light intensity. The response in each readout mode is fit with a line, and the 
ratio between the two slopes is the inter-calibration coefficient for one given
readout channel.
Ideally, this factor should be a simple constant between the two chip readout modes, 
but it turns out to depend on the SiPM signal form due to the different shaping times in the two modes. 
For longer SiPM signals (larger quenching resistor) the inter-calibration is bigger
than for shorter SiPM signals (smaller quenching resistor). 
The inter-calibration factors between the chip readout modes 
range between 4 and 13.

As with the gain, the inter-calibration extraction efficiency is
influenced by the quality of the LED light distribution system. The
inter-calibration coefficient extraction efficiencies during the 2007
and the 2008 data taking periods are plotted in Figure~\ref{gain_eff}
(right). 
After commissioning was completed, all channels with the exception of the 
2\,\% inactive channels and the channels connected to a broken LED, 
could be inter-calibrated. 
For the missing inter-calibration values the average of the module to which a SiPM belongs is used instead.


The uncertainty on the inter-calibration coefficient has been estimated 
from the comparison of several runs and is found to be better than 1\,\%. 
Temperature and voltage changes do not affect this coefficient 
since it is mainly driven by the stability of the components of the readout chip and of the
SiPM quenching resistor, all of which are stable in a range of 5--10 degrees.

\subsection{SiPM non-linearity}\label{sec:saturation}
Due to the limited number of pixels and the finite pixel recovery
time, the SiPM is an intrinsically non-linear device. The SiPMs used in
the AHCAL have a total of 1156 pixels with a recovery time between
25\,ns and 1\,$\mu$s, depending on the value of the quenching resistor. 

  The response function of a SiPM correlates the observed number of pixels 
  fired, $N_{\rm pix}$,  to the effective number of photoelectrons generated
, $N_{\rm pe}$, including cross-talk and after-pulses.
  The response of a SiPM can be approximated by the function 
 \begin{equation}
  \label{eq:sat}
  N_{\rm pix} = N_{\rm tot}\cdot(1-e^{-N_{\rm pe}/N_{\rm tot}}),
\end{equation}
  with $N_{\rm tot}$ the maximum number of fired pixels.
This formula is a useful approximation for the case of uniform light
distribution over the pixels and short light pulses.  

In the above approximation, one can extract a correction function for the SiPM non-linear response as the residual to linearity of the inverted SiPM response function, 
 \begin{equation}
  \label{eq:fsat}
 f_{\rm{sat}}(N_{\rm pix}) =  \frac{N_{\rm pe}}{N_{\rm pix}} = \frac{1}{N_{\rm pix}}\frac{log(1-N_{\rm pix}/N_{\rm tot})}{-1/N_{\rm tot}}.
\end{equation}

\begin{figure}[!t]
  \begin{center}
    \includegraphics[width=0.55\textwidth]{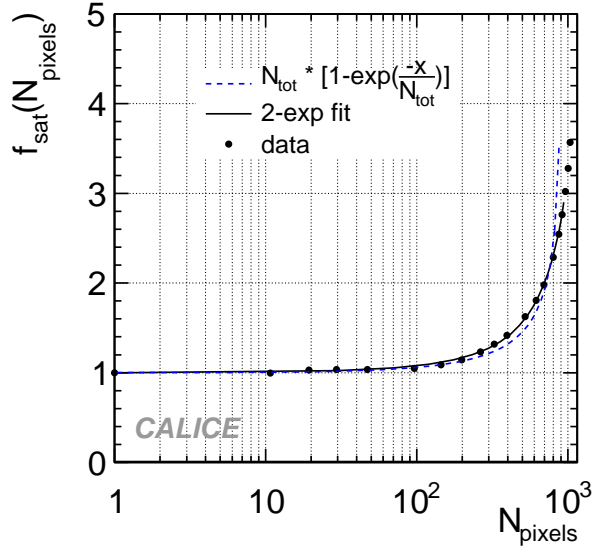}
  \end{center}
  \caption[]{{The SiPM non-linearity correction function, $f_{\rm{sat}}$. The points are the tabulated data of $N_{\rm pix}$ versus $N_{\rm pe}$ for one SiPM in the AHCAL. The solid line is obtained from the double exponential fit to the points, and it is used as a correction in this analysis. The dashed line is the single exponential approximation from equation~\ref{eq:fsat}.}}
  \label{sat}
\end{figure}

In the analysis presented here, we do not directly apply this formula but use tabulated measurements of $N_{\rm pix}$ versus $N_{\rm pe}$ for each SiPM instead. We fit the tabulated measurements with a double exponential function and use this to extract the value of $f_{\rm{sat}}(N_{\rm pix})$ applied in the calibration chain (Eq.~\ref{eq:calib}). 
The comparison of tabulated AHCAL data with the single and double exponential functions are shown in figure~\ref{sat} for one example SiPM. 

The use of a second exponential in the fitting function does not have a solid physics motivation yet, tough it could be explained thinking of two areas of the SiPM active surface differently illuminated from the WLS fiber. The total number of pixels is divided in two groups, and each of the group is described by an individual exponential function, such that the fitting function is the sum of two exponential functions like that in equation~\ref{eq:calib}. There is no reson for the groups to be exactly two; this method could be extended to more exponentials. For practical reasons though, one needs to limit the number of free parameters. The choice shown in figure~\ref{sat} turned out to be sufficiently accurate and stable. The correction factor is close to unity for signals of about 30\,pixels or 2\,MIPs, and increases exponentially up to infinity for signals in saturation.


As illustrated for one SiPM in figure~\ref{sat}, 
the response curves of each SiPM has been sampled with 20 measurement points on a test bench
setup illuminating each SiPM with LED light of variable intensity. For these studies, the SiPMs were not mounted on tiles, but were bare SiPMs. Therefore, all
the pixels have been illuminated with light in a homogeneous way. The
 measurement results for all SiPMs installed in the AHCAL are
given in~\cite{ahcal}. The maximum number of fired
pixels ($N_{\rm tot}$(bare)) for each SiPM is extracted with a fit to the measured points using Eq.\ref{eq:sat}.
The spread (RMS) in the values of $N_{\rm tot}(bare)$ between all the curves is about 20\,\%. SiPMs with 
$N_{\rm tot}(bare)>$~900 have been pre-selected.
This ensures not too large variations in the non-linear response
function of each device. 
The 20 measurement points for each SiPMs are stored in a database. 
A linear interpolation of these points is used to calculate $f_{\rm{sat}}$ from Eq.\ref{eq:fsat} and
linearize the calorimeter response during data reconstruction.

\begin{figure}[!b]
  \begin{center}
    \includegraphics[width=0.5\textwidth]{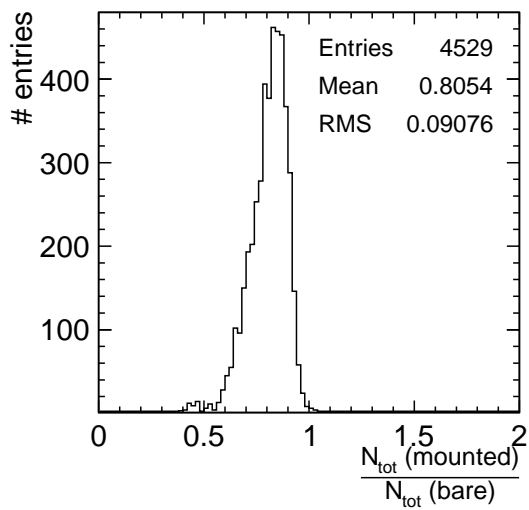}
  \end{center}
  \caption[]{{Ratio of maximum number of fired pixels, $N_{\rm tot}$(mounted),
      measured with SiPM mounted on a tile to $N_{\rm tot}$(bare) measured 
      directly with bare SiPMs. }}
  \label{fig:saturation}
\end{figure}

Alternatively, $N_{\rm{tot}}$ has also been extracted using the AHCAL LED monitoring system 
from measurements with the with SiPM mounted on a tile ($N_{\rm tot}$(mounted)). 
As the saturation point in number of pixels is independent on the linearity of the light, 
no correction of the LED light intensity with PIN diode has been applied for this study.
Figure~\ref{fig:saturation} shows the ratio of $N_{\rm tot}$(mounted) to $N_{\rm tot}$(bare). 
The plot shows that the maximum number of pixels in the in-situ setup is on average 80.5\,\% of the
value determined in the laboratory setup~\cite{nils_diploma} with a
wide distribution (RMS=9\,\%). This factor is interpreted as geometric mismatch between
the WLS fiber and the photodetector.  
The fiber has a  1\,mm diameter while the SiPM active surface area is 1$\times$1\,mm$^2$; 
the geometric ratio between areas is 79\,\%, in agreement with the measured value. 
Therefore, only a fraction of the SiPM surface is illuminated and the 
laboratory curves are re-scaled by the measured value of 80.5\,\% to
correct for this effect before they are used to correct for the SiPM
saturation. 

The uncertainty of the determination of the saturation point for a single 
channel is lower than 3\,\%, if the LED light range properly covers the 
SiPM saturation region, and if this region is measured well below the 
ADC saturation. Unfortunately, these conditions are true only for a sub-sample 
of about 73\,\% channels. 
Also judging from the tails in the distribution of Figure~\ref{fig:saturation} 
some of the fitted results need to be investigated more accurately.
For this reason, an average scaling factor is used for all channels.
Further studies will address the possibility of using a channel-by-channel factor instead.
Furthermore, the measured SiPM response points, from which the correction of non-linear detector response is 
calculated, are affected by the SiPM gain uncertainty of 2\,\%, discussed in the following section.


\section{The test beam experiment}\label{sec:testbeam}
\subsection{The experimental setup at CERN}\label{sec:beamline}
\begin{figure}[!t]
  \begin{center}
    \includegraphics[width=0.99\textwidth]{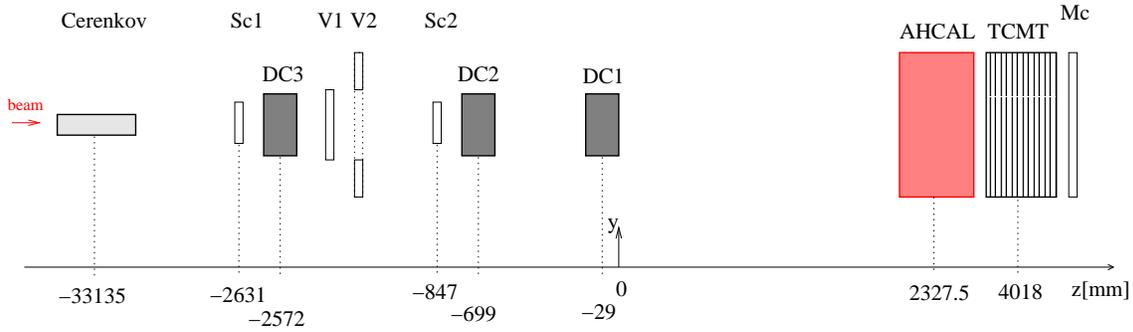}
  \end{center}
  \caption[]{{ Top view of the CERN beam test setup. The plot shows the
      instrumentation in 2007 (the $y$-axis is not to scale). The beam enters from the left side. See text for explanations of the components. }}
  \label{fig:beamline}
\end{figure}

The data discussed in the following were collected in July 2007
at the CERN  SPS test beam facility H6. A sketch of the experimental setup is shown in
Figure~\ref{fig:beamline}. Apart from the fully equipped AHCAL and a
prototype of a tail-catcher and muon tracker (TCMT~\cite{TCMT}), the
beam installation consists of various trigger and beam monitoring
devices. A threshold \v{C}erenkov counter was used to discriminate
between electrons and pions. The beam trigger was defined by the
coincidence signal of two plastic scintillator counters with
$10\times10$\,cm$^2$ area, referred to as Sc1 and Sc2 in
Figure~\ref{fig:beamline}. One scintillator trigger (V1), with an area
of $20\times20$\,cm$^2$ and analog read out, tagged
multi-particle events. Another scintillator with a
$100\times100$\,cm$^2$ surface and a $20\times20$\,cm$^2$ hole in the
center (V2), was used to reject the beam halo. Three delay wire chambers
(DC1, DC2 and DC3) were used to monitor the beam and reconstruct
tracks. Events tagged by a scintillator with
$100\times100$\,cm$^2$ area (Mc1), placed behind the TCMT are most likely to be muons.

During most of the tests, a silicon tungsten
electromagnetic calorimeter~\cite{SiW-ECAL} was placed in front of the
AHCAL, but this was not the case for the results reported here. 
The AHCAL was placed on a movable stage, which could
shift the detector vertically and horizontally. In addition, the detector can 
be rotated with respect to the beam direction
from an angle of $90^{\circ}$ (beam normal to the detector plane) 
to approximately $60^{\circ}$.

\subsection{Monte Carlo simulation}\label{sec:simulation}
The test beam setup as shown in Figure~\ref{fig:beamline} 
is simulated with Mokka~\cite{Mokka}, a {\sc Geant}4-based~\cite{G4} Monte Carlo program, 
followed by a
digitization package simulating the response of the detector and electronics. 
The particle source of the simulation is positioned upstream of the \v{C}erenkov detector.
The beam position and spread are chosen to match the beam shapes measured in data by the 
delay wire chamber, DC3. The beam particles are parallel to the beam axis, according to the 
measurements in the three delay wire chamber detectors.
The material upstream of the AHCAL is simulated. The sub-detectors are simulated with different levels of
detail, depending on their impact on the physics analysis: material
simulation only for the \v{C}erenkov counter, raw energy depositions
stored for the trigger counters, and partial  electronics simulation for
the tracking detectors. For the AHCAL, the simulation gives the
raw energy depositions in a virtual scintillator grid of
$1\times1$\,cm$^2$ tile size.
The simulation is followed by a digitization procedure, which takes
into account
\begin{itemize}
\item the realistic detector granularity,
\item light cross-talk between neighboring tiles,
\item non-linearity and statistical fluctuations on the pixel scale,
\item SiPM and readout electronics noise.
\end{itemize}
The actual geometry of the AHCAL is simulated by summing up the signal yield of 9 (36, 144) virtual cells to obtain those of the actual geometry $3\times3~(6\times6,~12\times12)\,\textrm{cm}^2$ cells.

Light cross-talk between neighboring cells, due to the imperfect
reflective coating of the tile edges, is simulated assuming that
from each 3~cm-long tile edge 2.5\,\% of the scintillator light leaks homogeneously
to the neighboring tile. This value is scaled to take into account the fraction of edge shared
with the neighbors for cells of different size.
The amount of light cross-talk was checked experimentally only for two tiles.
The leakage from one tile edge was quantified to be about 2.5\,\%. 
No information on the spread of this value between all tiles is given.
This value is expected to influence the energy reconstructed and 
the transverse shower profile.
From the comparison of the  energy reconstructed in simulation and data, 
the value of 2.5\,\% for the light cross-talk on each tile edge is found to be adequate. 
A light cross-talk of 1.25\,\% or 3.75\,\%
leads to a difference in the energy scale between data and Monte Carlo larger then 5\,\%. 

To simulate the non-linear behavior of the photodetectors, the energy
deposition is translated from GeV to the number of fired SiPM pixels.
For this, an intermediate step converts the response simulated
in units of GeV to MIP equivalents. The conversion factor is estimated
from the simulation of an 80\,GeV muon beam in the AHCAL and is found
to be 816\,keV/MIP, corresponding to the energy lost by a minimum ionizing particle in the scintillator. The
amplitude in units of MIPs is then converted into pixels, using the
measured light yield for each individual channel. With this scale the
measured SiPM response curves from the test bench are used to simulate
the SiPM non-linearity. Where not available, the curve of the next neighboring 
tile is applied.

If $N_{\rm{pix}}$ is the amplitude in pixels obtained this way and
$N_{\rm{max}}$ is the saturation level of the individual channel,
statistical effects are accounted for by generating a binomial random
number with $N_{\rm{max}}$ repetitions and a probability of
$N_{\rm{pix}}/N_{\rm{max}}$. The result is treated as the
number of pixels firing for this specific event, and is translated
back to the MIP scale with the channel-specific light yield.

At this stage, the Monte Carlo signal simulated the response of
the AHCAL to the energy deposited by particles in an event. However,
both the electronic components and the SiPM dark current induce noise.
This noise component is assumed to be completely independent of the
physics signal in each channel. The noise distribution is 
non-Gaussian due to the positive SiPM dark noise component. As the 
frequency of dark noise and the amount of inter-pixels crosstalk varies 
from SiPM to SiPM, the best way to account for noise is to
take it from the data rather than try to simulate it. 
Noise events for each calorimeter cell are taken from data, 
are calibrated to the MIP scale according to the reconstruction procedure 
described in section~\ref{sec:calibration}, and are added randomly 
to the deposited energy of that given cell of a simulated event.


A cell that could not be calibrated in the real detector,
either due to an inactive photodetector or to missing calibration values, is
also ignored in the simulation. This is about 2\,\% of the total number 
of cells in the calorimeter.


\section{Calorimeter response to positrons}\label{sec:EMresponse}
\subsection{Selection of positron events}\label{sec:selection}
The analysis presented here is based on positron runs between 10 and 50\,GeV.
Each energy point has more than 150k recorded beam triggers. All
positron runs have been simulated with statistics similar to the
corresponding data runs.

Single positron showers are selected for analysis using the beam instrumentation.
Although the beam configurations are set to deliver a positron
enriched beam, some contamination, mainly from muons, exists. 
The pion contamination is expected to be negligible, since the tertiary positron beam is produced from a higher-energy mixed beam impinging on a thin (2 $X_0$) lead target which does not result in the production of lower-energy tertiary pions. 
The muon contamination originates from in-flight decay of hadrons upstream of the production target, which results in a muon component that passes the momentum selection. 

Cells with a signal above threshold are called hits and 
$E_{\rm{hit}} > 0.5$\,MIP is required.
To reject empty events that can occur due to random triggers or scattered
  particles, the number of hits has to be $N_{\rm{hit}} > 65$.
Furthermore, the energy weighted center-of-gravity in the beam direction ($z$), defined as
$\langle {z} \rangle = \sum_{i} z_i E_i/\sum_{i} E_i$,
has to be $\langle {z} \rangle < 390$\,mm (about half of the calorimeter depth). 
This requirement eliminates muons, which
deposit their energy equally distributed over the entire calorimeter
depth, as opposed to electrons which have a short shower contained in the first half of the calorimeter. It was found that this muon rejection was more efficient than the selection based either on the  \v{C}erenkov counter, which does not provide electron-muon separation for 30 GeV and above, or on the muon trigger Mc1 which has an efficiency of about 50\%. 
Particles which interact in the material upstream of the AHCAL are removed by requiring a 
good track in the delay wire chambers ($\chi^2 /{\rm dof} < $ 6), and a MIP-like energy deposition in the multiplicity counter (V1).
With these selection criteria, 45\,\% of all recorded events at
10\,GeV are accepted. 
According to Monte Carlo studies
99.9\,\% of all electron events pass the
selection criteria, whereas 99.8\,\% of all muon events are rejected. The typical fraction of muons in a run is about 5-10\,\%.

The uncertainty on the mean energy of the beam
 is reported in~\cite{ECAL_emPaper} to be
\begin{equation}
  \frac{\Delta E_{\rm{beam}}}{E_{\rm{beam}}} 
  = \frac{0.12}{E_{\rm{beam}}\,[\rm{GeV}]}
  \oplus 0.1\,\%.
\end{equation}
The first term is related to hysteresis in the
bending magnets while the calibration and the uncertainties on the
collimator geometry give the constant term. Since this uncertainty is
negligible compared to the detector uncertainties, we assume the beam
energy to be fixed.
The dispersion of the beam energy can be calculated according to~\cite{H6manual}
from the settings of the momentum selecting collimators on the 
beam line and is below 0.24\,\%
for all the runs in this analysis.

\subsection{Linearity}\label{sec:linearity}

\begin{figure}[!t]
  \begin{center}
    \includegraphics[width=0.6\textwidth]{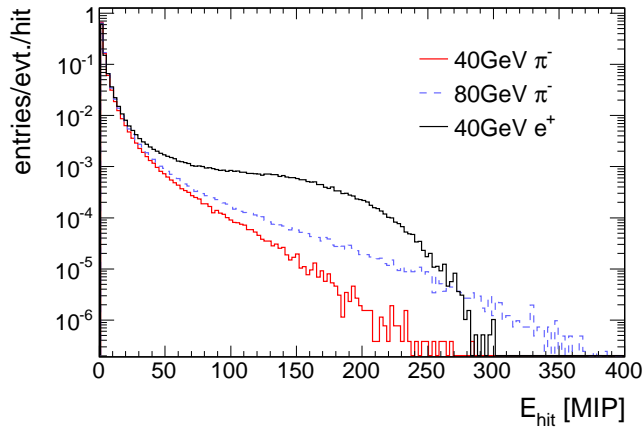}
  \end{center}
  \caption[]{{ Hit energy spectrum for 40\,GeV positron showers 
       compared to that of 40\,GeV  and 80\,GeV pion showers  from a {\sc Geant}4 simulation.  }}
  \label{fig:hitenergy}
\end{figure}

The linearity of the  calorimeter response for a large range of incident particle energies is a key feature, which allows for an important test of the calibration chain.
Electromagnetic showers offer the most rigorous test for non-linearity correction, 
since the energy deposited per single tile in an electromagnetic
shower is larger than that in a hadronic shower for the same particle energy.
Figure~\ref{fig:hitenergy} shows the hit energy spectrum of a 40\,GeV
positron shower compared to the spectra of 40\,GeV and 80\,GeV pion showers. The positron shower 
clearly has more hits with high
energy deposition, even when the total particle energy is only half that of the pion.

\begin{figure}[!t]
  \begin{center}
    \includegraphics[height=0.37\textwidth]{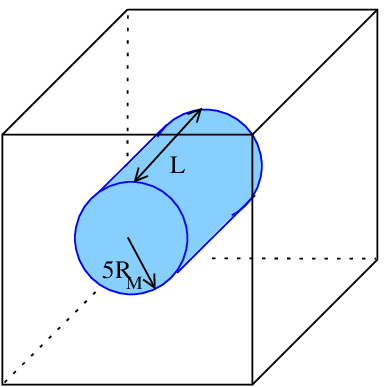}
    \includegraphics[width=0.58\textwidth]{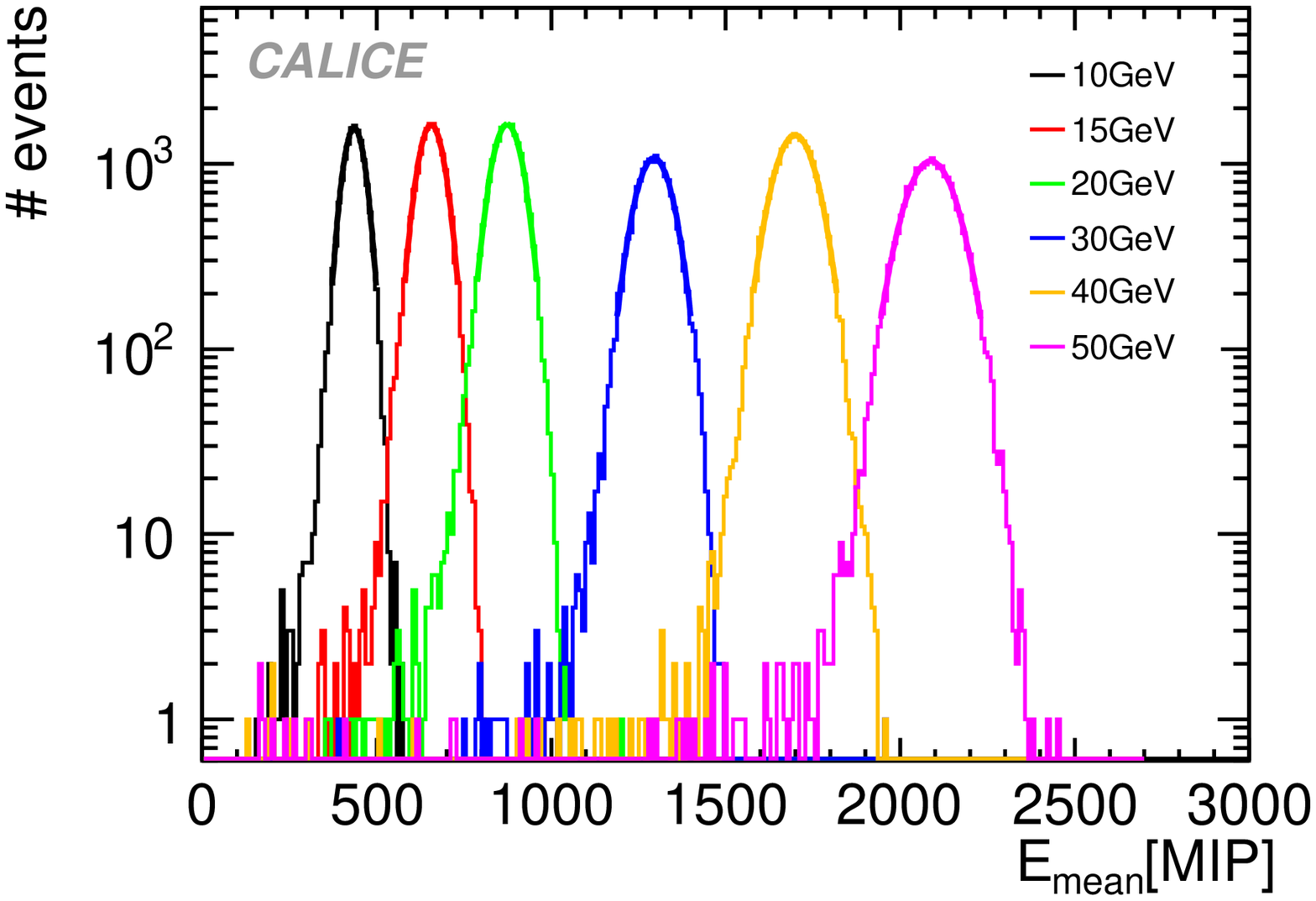}
  \end{center}
  \caption[]{{ The shower energy is summed up in a cylinder (left); see text for details. Spectra of the energy sum for positron data with energy between 10 GeV and 45 GeV (right). For each spectrum  the mean energy response in units of MIP, $E_{\rm{mean}}$, is obtained with a Gaussian fit in the range $\pm 2\sigma$.}}
  \label{fig:spectra}
\end{figure}

A set of positron runs with incidence normal to the center of the calorimeter face is analyzed.
To minimize the influence of noise, the energy is summed up in a
cylinder around the shower axis, where the shower axis is defined by projecting a track formed in the tracking system into the first layer of the AHCAL. This cylinder, sketched
in Figure~\ref{fig:spectra} (left), has a radius of 5 Moli\`ere radii
($r=5~R_{\rm M}$, with $R_{\rm M}=2.47$\,cm~\cite{ahcal}), which ensures a lateral containment of more than 99\,\% of
the shower energy. 

\begin{figure}[!t]
  \begin{center}
    \includegraphics[height=0.6\textwidth]{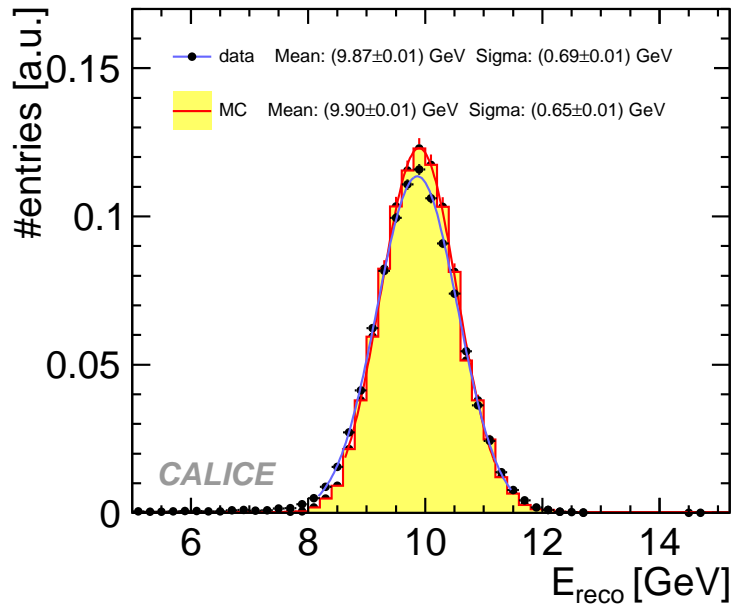}
  \end{center}
  \caption[]{{
      Reconstructed energy of a 10\,GeV positrons for data (dots) and for Monte Carlo (filled histogram), as
      well as a Gaussian fit to data (blue line).}}
  \label{fig:eSum}
\end{figure}

The length $L$ of the cylinder is chosen to contain
the whole shower energy. 
As suggest by simulations of 50 GeV electron showers, setting L to 20 layers
contains the showers.
Figure~\ref{fig:spectra} (right) shows the final reconstructed spectra for positron runs in the energy range 10 to 50 GeV. The positrons are normally incident on the calorimeter front face, with a distribution centered in the same calorimeter cell for each run. 
The distribution is fit with a Gaussian function in the range $\pm 2\sigma$. 
The position of the peak is taken as the mean energy response, $E_{\rm{mean}}$, measured in units of MIP.
The reconstructed energy of a 10\,GeV positron shower is compared to the digitized energy from a Monte Carlo simulation in Figure~\ref{fig:eSum}. 
The agreement between data and simulation is satisfactory. 

The statistical uncertainties on the mean energy deposition are
negligible. The main source of systematic uncertainties is
2\,\% from MIP scale calibration. 
The uncertainty of 2\,\% on the SiPM gain determination, 
resulting from the fit stability and the uncertainty on the 
determination of the SiPM saturation level both affect the correction of the SiPM non-linear response.
For the saturation level a common re-scaling factor is applied to all SiPM curves
determined in the laboratory setup. The rescaling is needed to account for the
partial illumination of the SiPMs from the WLS fiber as discussed in Section \ref{sec:saturation}. 
As shown in
Figure~\ref{fig:saturation}, the ratio between the in-situ measured SiPM
saturation level and the test-bench determined value has a wide
distribution. 
Since a common factor of 80.5\,\% is used to rescale all SiPM response curves, 
an uncertainty of 11.3\,\% on this value is assumed, which represents the 
spread of all measured values as taken from Figure~\ref{fig:saturation}. 
To account for this uncertainty in an uncorrelated 
way for all SiPMs, 100 experiments have been performed assigning different rescaling 
coefficients for each channel, generated randomly with a Gaussian distribution centered at 0.80 and 
with a sigma of 0.09. For each experiment the energy in the calorimeter is 
reconstructed, using the set of curves rescaled by these randomly generated coefficients 
to correct the non-linear SiPM response. Finally, the 
one standard deviation spread of the 100 reconstructed energies from these simulated experiments is taken as 
the systematic uncertainty for the reconstructed energy.
All of the above listed systematic
uncertainties are uncorrelated and thus added in quadrature. The total systematic
uncertainty ranges from $0.2~\rm{GeV}$ (2\,\%) at 10 GeV to 1.7 GeV (3.4\,\%) at 50 GeV.

The reconstructed energy in GeV is obtained as $E_{\rm{reco}} = E_{\rm{mean}}/w$, where $w$ is the electromagnetic energy scale factor (MIP-to-GeV).  The scale factor is determined with a linear fit 
from zero to 50\,GeV to the distribution $E_{\rm{mean}} [{\rm MIP}]$ versus $E_{\rm{beam}} [{\rm GeV}]$.
The resulting values for data and Monte Carlo are $w_{\rm{data}}= (42.3 \pm 0.4)$\, MIP/GeV and $w_{\rm{MC}} = (42.0 \pm 0.4)$\,MIP/GeV, respectively. Within the uncertainties, the scale factors are in good agreement.
\begin{table}[!b]
  \centering
  \begin{tabular}{l|c c c c c|c c}
        & \multicolumn{5}{c|}{Data}         & \multicolumn{2}{c}{MC} \\
\hline
$E_{\rm beam}$ &$E_{\rm reco}$ &$\delta_E^{\rm MIP}$ [\%]&$\delta_E^{\rm Gain}$ [\%]&$\delta_E^{\rm sat}$ [\%]&$\Delta_E^{\rm tot}$ [GeV] & $E_{\rm reco}$ & $\Delta_E^{\rm tot}$ [GeV]  \\ \hline
10      & 9.9    &2.0&0.3&0.4& 0.2 &             9.9    &0.2\\
15      &15.0    &2.0&0.5&0.8& 0.3 &            15.0    &0.3\\
20      &20.1    &2.0&0.7&1.2& 0.5 &            20.2    &0.5\\
30      &29.9    &2.0&1.1&1.8& 0.9 &            30.4    &0.9\\
40      &39.3    &2.0&1.2&2.3& 1.3 &            40.8    &1.3\\
50      &48.3    &2.0&1.4&2.6& 1.7 &            51.0    &1.8\\
\hline
  \end{tabular}
  \caption{AHCAL energy reconstructed in data and MC (in units of GeV) for various positron beam energies. The table reports the values plotted in Figure~10. The systematic uncertainties for data are detailed in their percentage values. 
The total absolute error $\Delta_E^{\rm tot}$ is the sum in quadrature of 
the uncertainties on the MIP, on the SiPM gain and on the saturation point determination.}
  \label{tab:linearity}
\end{table}

\begin{figure}[!t]
  \begin{center}
    \includegraphics[width=0.6\textwidth]{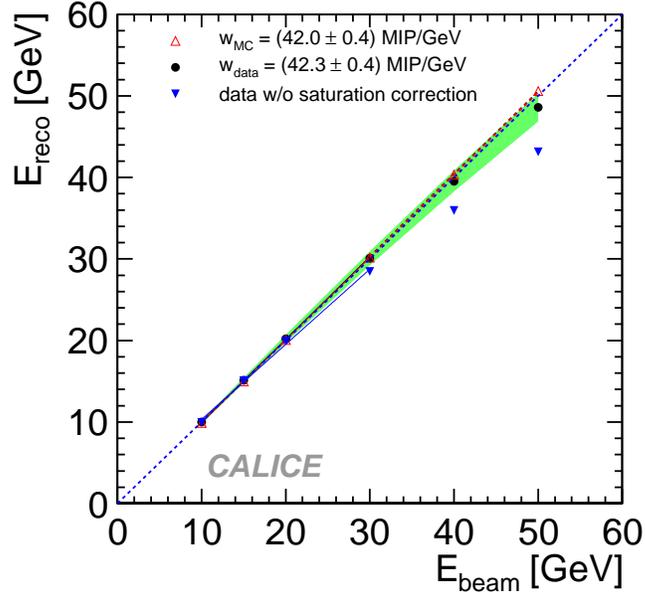}
  \end{center}
  \caption[]{{Linearity of the AHCAL response to positrons in the
      range 10--50\,GeV. The blue dotted line shows exact linearity. Dots correspond to data corrected for SiPM non-linear response, 
      blue triangles show the data before this correction, and the open triangles show the simulation.
      The green band indicates the systematic uncertainty as quoted in Table~1, $\Delta_E^{\rm tot}$ [GeV].}}
  \label{em_linearity}
\end{figure}

The linearity of the AHCAL response to positrons is shown in Figure~\ref{em_linearity}. A comparison of the data before and after correction for the SiPM non-linear response indicates the magnitude of this correction, which does not exceed 10\,\% even at 50 GeV positron energy.
The values shown in Figure~\ref{em_linearity} are reported in Table~\ref{tab:linearity}.

\begin{figure}[!t]
  \begin{center}
    \includegraphics[width=0.49\textwidth]{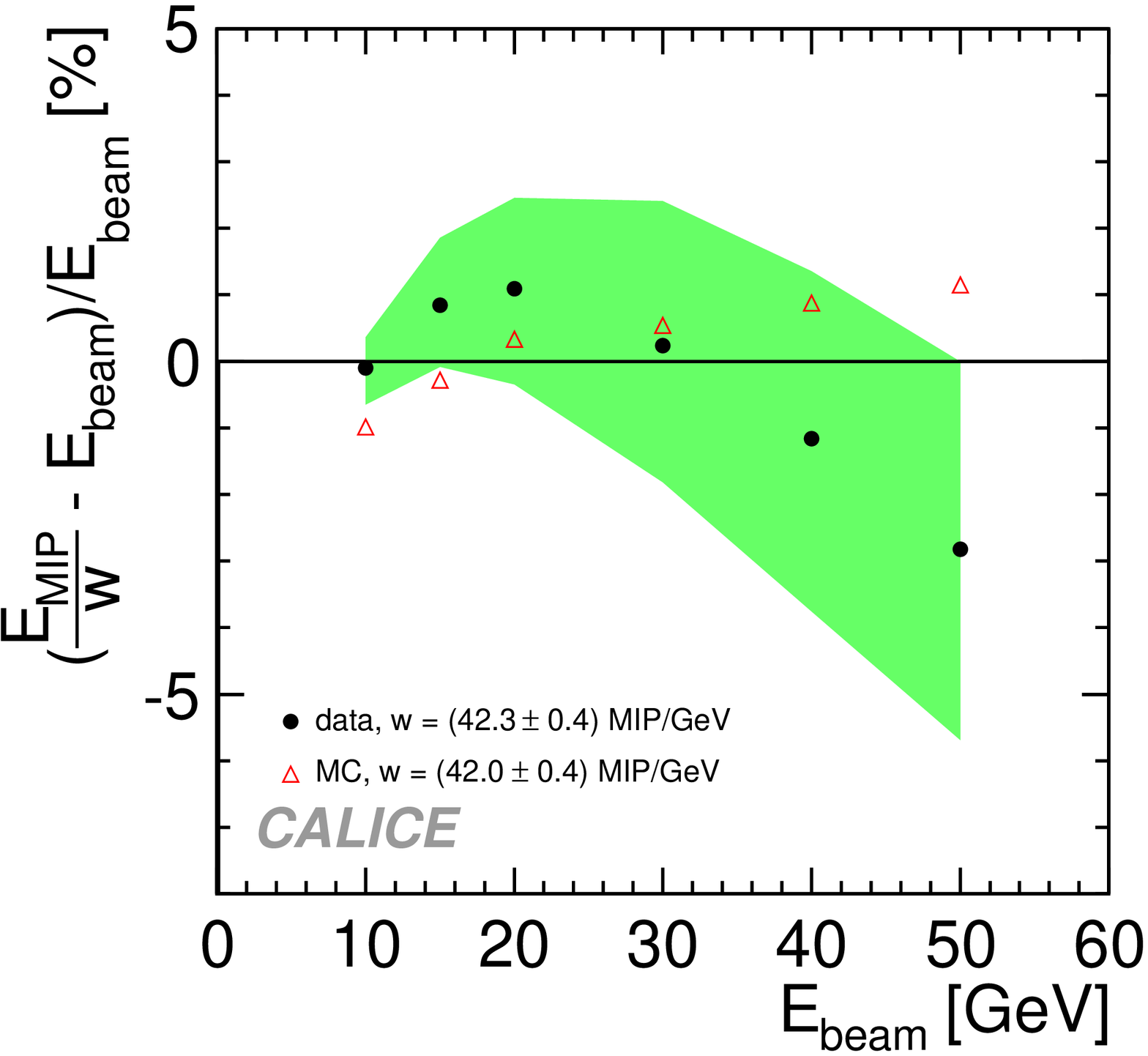}
    \includegraphics[width=0.49\textwidth]{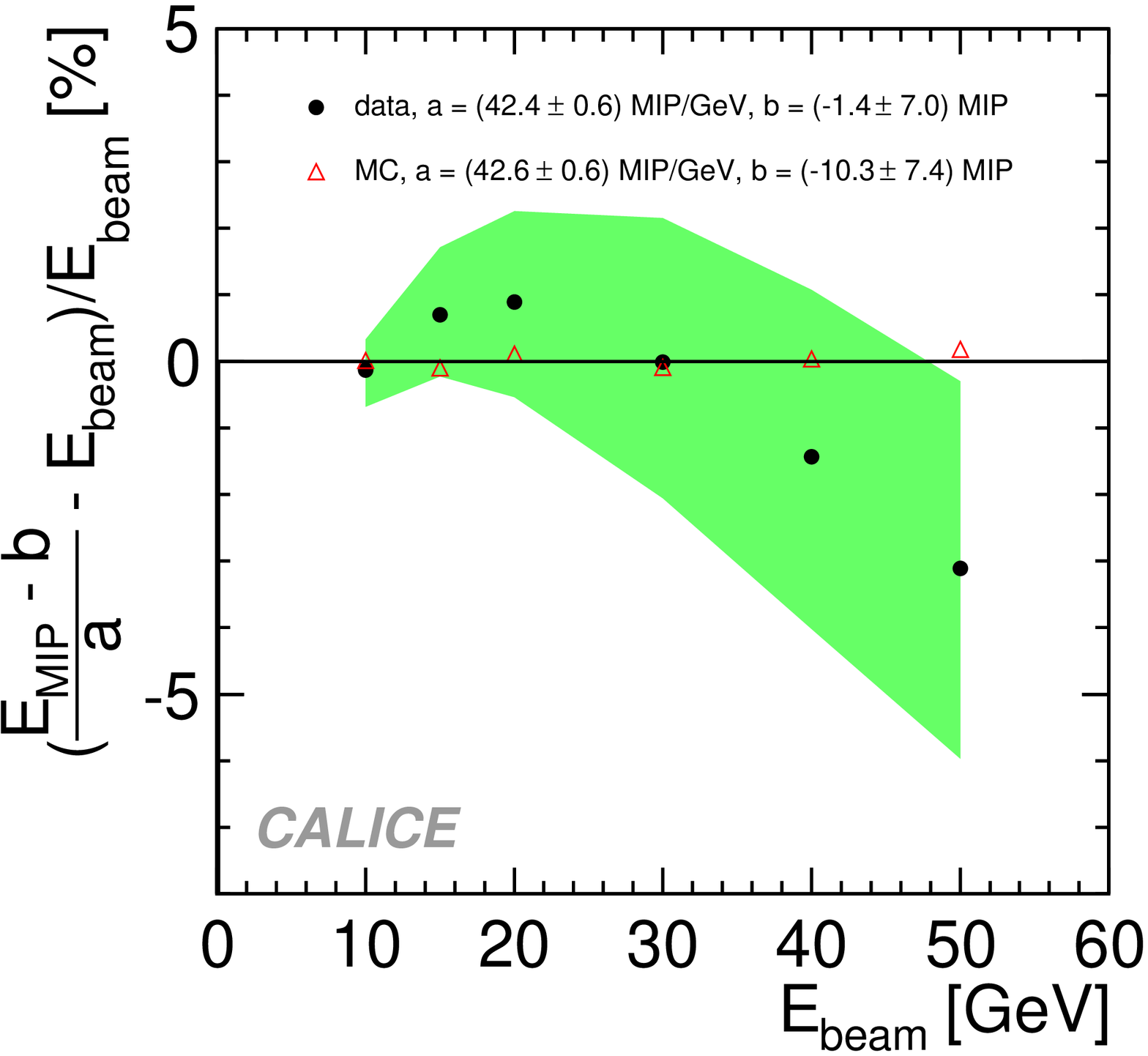}
  \end{center}
  \caption[]{{ Residual to a fit of the data and Monte Carlo points presented in Figure~\ref{em_linearity} using, the function, $y=ax$, 
      (left), and the function, $y=ax+b$ (right), in the range 10--50\,GeV. 
      Dots correspond to data, and open triangles to
      simulation. The green band indicates the sum in quadrature of the energy dependent systematic uncertainties, $\delta_E^{\rm Gain}$ and $\delta_E^{\rm sat}$ in Table~1.}}
  \label{em_residual}
\end{figure}

The residuals for data and Monte Carlo are presented in Figure~\ref{em_residual}. 
Here, the green band indicates the quadratic sum of the energy dependent systematic uncertainties. In Table~1
the contribution to the uncertainty from the SiPM gain variation, $\delta_E^{\rm Gain}$, and from the saturation point determination, $\delta_E^{\rm sat}$ are listed. 
The uncertainty on the MIP scale, $\delta_E^{\rm MIP}$, cancels in the ratio since the same calibration constants are used in data and Monte Carlo. 
In Figure~\ref{em_residual} (left), the residuals from the linear function 
suggests a non-zero offset at zero energy.  This negative offset is the combined effect of the 0.5 MIP threshold (loss of energy) and the detector noise (addition of energy).
Instead of the more conventional linear function with $b=0$, the function
$E_{\rm{mean}} = a\cdot E_{\rm{beam}} + b $ can be used to fit the data in
the range 10--50 GeV. A value of $b=-10.3\pm 7.4$\,MeV is found for the Monte
Carlo offset. Once this offset is removed the Monte Carlo linearity is better than 0.5\,\% over the whole range, as shown in the right plot of Figure~\ref{em_residual}.

The deviation from linearity (Fig.~\ref{em_residual} left) in data is less than 1\,\% in the
range 10 to 30\,GeV and the maximum deviation is about 3\,\% at
50\,GeV. The remaining non-linearity at high energies hints at problems
with the rescaling of the saturation curves, as described in
Section~\ref{sec:saturation}. This behavior is not sufficiently
reproduced in the Monte Carlo digitization, where the same curve is
used to simulate saturation as is used to correct for it. 

\begin{figure}[!t]
  \begin{center}
    \includegraphics[width=0.48\textwidth]{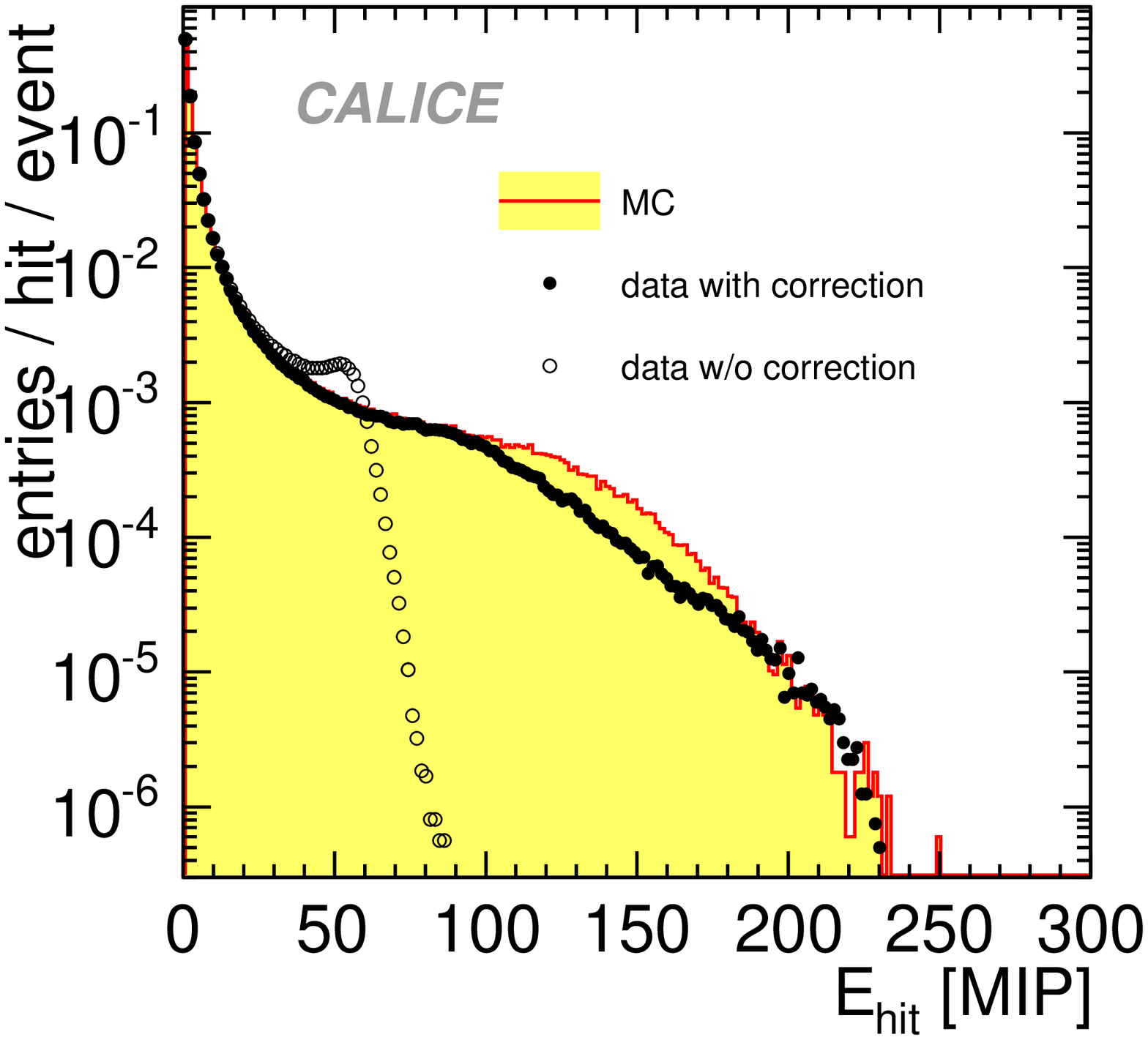}
    \includegraphics[width=0.48\textwidth]{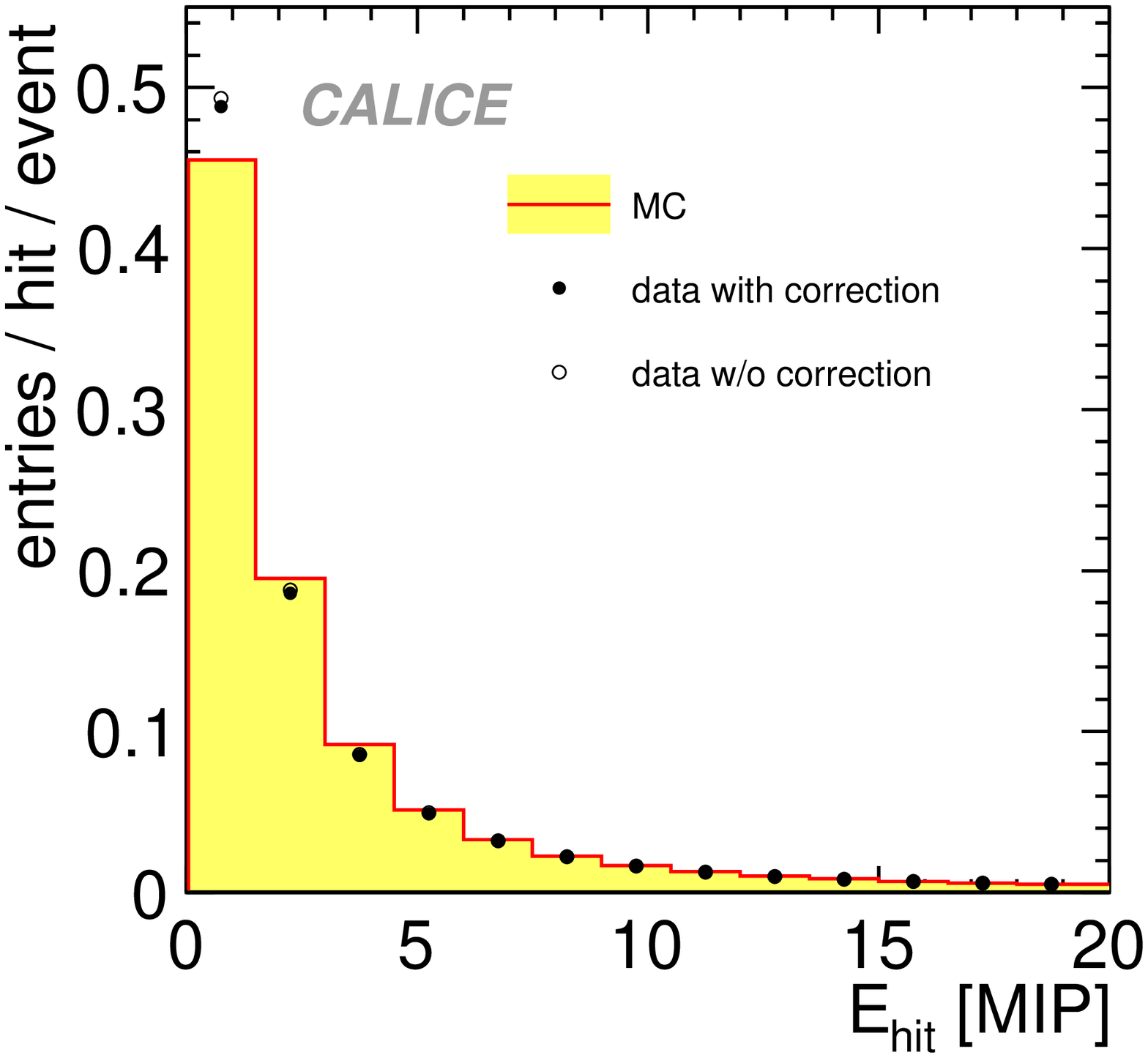}
  \end{center}
  \caption[]{{ Hit energy spectrum for 30\,GeV positron showers in the
      AHCAL. Open circles (dots) show the data before (after) 
      correction for the non-linear response of the SiPM.
      The left plot shows the hit distribution in a logarithmic scale and the right plot on a linear scale. The shaded histogram is from digitized simulation.}}
  \label{em_hitenergy}
\end{figure}

The impact of the saturation correction is better seen in
Figure~\ref{em_hitenergy} where the energy per hit is shown with
and without the correction factor $f_{\rm{sat}}$ applied, for
 30\,GeV electromagnetic showers. Whereas the correction is negligible for low signal amplitudes, 
it becomes significant at larger
amplitudes, resulting in a strong correction for the tail of the
distribution. The maximum energy deposited in one cell for a 30\,GeV electromagnetic shower is
$\sim$230 MIPs corresponding to about 3450 pixels (assuming a light yield, LY =
15 pixel/MIP). For this amplitude the correction factor is
$f_{\rm{sat}}(A_i) \sim 3.1$. The remaining miss-match between data and Monte Carlo around
  100-200 MIPs is an effect of the non-perfect correction of the non-linear SiPM response.
This imperfect correction affects only a small fraction of the total energy; the hits above 50 MIPs contribute only 0.5\,\% (4\,\%) of the total energy at 10 GeV (40 GeV). 

\subsection{Electromagnetic energy resolution}\label{sec:resolution}
\begin{figure}[!t]
  \begin{center}
    \includegraphics[width=0.7\textwidth]{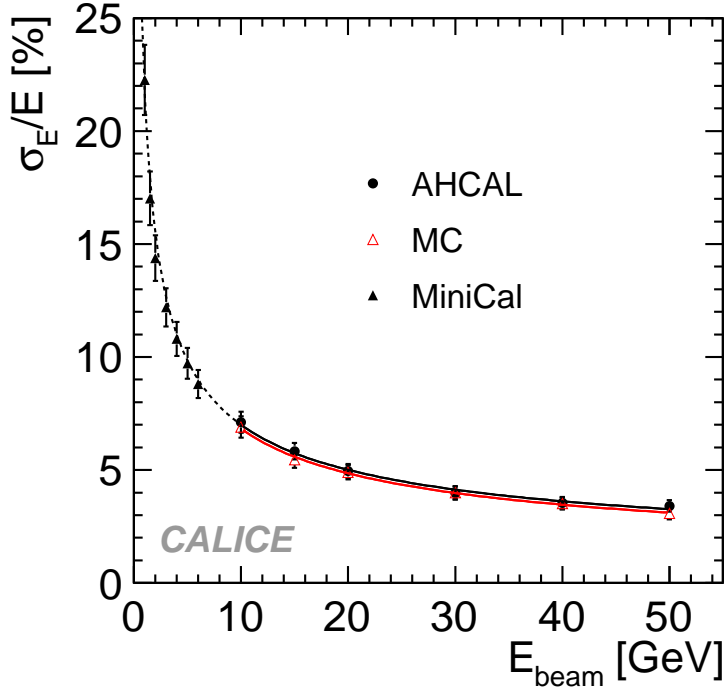}
  \end{center}
  \caption[]{{ Energy resolution of the AHCAL for positrons (dots). The resolution agrees with that of a previous prototype
      (full triangles) with the same sampling structure. The errors are the quadratic sum of statistics and systematic uncertainties. The open triangles are the obtained from the analysis of the digitized simulated events. Fit curves to the data and MC are shown in the region 10--50 GeV. The dashed line is the extrapolation of the fit to AHCAL data in the low energy region covered by the MiniCal data. }}
  \label{fig:em_Eres}
\end{figure}

Energy resolution is a principal figure of merit in calorimetry and is
estimated as the width divided by the mean of a Gaussian fit to the energy
sum within $\pm\,2\,\sigma$ of the mean of an initial fit over the full range. 
The resolution achieved with the AHCAL is plotted as a function
of the beam energy in Figure~\ref{fig:em_Eres}. 
The values shown in this figure are reported in Table~\ref{tab:resolution}.
Fitting the AHCAL energy resolution in a range of $\pm 2\sigma$, with
\begin{equation}
  \frac{\sigma_E}{E} = \frac{a}{\sqrt{E}} 
  \oplus b \oplus \frac{c}{E} 
\end{equation}
results in a stochastic term of $a =(21.9 \pm 1.4)\,\% \sqrt{E\,[\rm{GeV}]}$, whereas the
constant term is $b = (1.0 \pm 1.0)\,\%$. The noise term of $c =
58.0$\,MeV is extracted from the spread (RMS) of the 
random trigger event distribution and kept constant during the fitting procedure. 
The energy resolution agrees well with that of an earlier
prototype (Minical) with 108 channels and of the same sampling~\cite{MiniCal}, that
was tested in the energy range between 1 and 6\,GeV and reached a
resolution with a stochastic term $a =(20.7 \pm 0.7)\,\% \sqrt{E\,[\rm{GeV}]}$ 
and a constant term $b =(2.6 \pm 1.3)\,\%$.

The energy resolution of the simulation is found to have a
stochastic term of $a =(21.5 \pm 1.4)\,\% \sqrt{E\,[\rm{GeV}]}$, a constant term of $b =
(0.7 \pm 1.5)\,\%$ and again a fixed noise term of $c = 58.0$\,MeV.
Within the fit uncertainty, the stochastic terms of data and
simulation are in good agreement.
The noise term is fixed to the same value as for data 
since the noise in the simulation is artificially added from random trigger data events.
The constant term $b$, representing calibration uncertainties and non-linearities, is consistent with
zero in the simulation as expected, since the same curves are both in the simulation of the non-linear SiPM response and in its correction.

\begin{table}[htbp]
  \centering
  \begin{tabular}{l|c c|c c}
        & \multicolumn{2}{c|}{Data}         & \multicolumn{2}{c}{MC} \\
\hline
$E_{\rm beam}$ [GeV] & $\sigma_E/E$ [\%]&Uncertainty [\%]&  $\sigma_E/E$ [\%]&Uncertainty [\%]\\
\hline
10    &  7.11 &   0.47  &          6.90 &   0.49\\
15    &  5.83 &   0.36  &          5.45 &   0.38\\
20    &  4.95 &   0.32  &          4.90 &   0.34\\
30    &  3.97 &   0.29  &          4.00 &   0.31\\
40    &  3.54 &   0.26  &          3.51 &   0.27\\
50    &  3.41 &   0.25  &          3.07 &   0.26\\

\hline
  \end{tabular}
  \caption{AHCAL energy resolution in data and MC for various positron beam energies. The table reports the values plotted in Figure 13. 
The listed uncertainties include statistical uncertainties and systematic uncertainties added in quadrature.
}
  \label{tab:resolution}
\end{table}

\subsection{Shower profiles}\label{sec:profiles}
The longitudinal profile of a shower induced by a particle with
incident energy $E$ in GeV traversing a matter depth $t$ can be
described as~\cite{Leroy}
\begin{equation}\label{eq:profile}
  f(t) = \frac{\,dE}{\,dt} = at^{\omega} \cdot e^{-bt},
\end{equation}
where the parameter $a$ is an overall normalization, and the parameters $\omega$ and $b$ are 
energy and material-dependent. The first term represents the fast
shower rise, in which particle multiplication is ongoing, and the
second term parametrizes the exponential shower decay.
Given this parametrization with $t$ in units of radiation lengths,
the particle multiplication and the energy deposition reach their
maximum after
\begin{equation}\label{eq:t_max}
  t_{\rm{max}} =  \left[\ln{\frac{E}{\epsilon_{\rm c}}}-0.5\right]
\end{equation}
radiation lengths from the beginning of the cascade of a particle with energy $E$. 
The critical energy, $\epsilon_{\rm c}$ is a property of the calorimeter material and does not 
depend of the energy of the particle.
The position $t_{\rm max}$ is called the shower maximum.

\begin{figure}[!t]
  \begin{center}
    \includegraphics[width=0.44\textwidth]{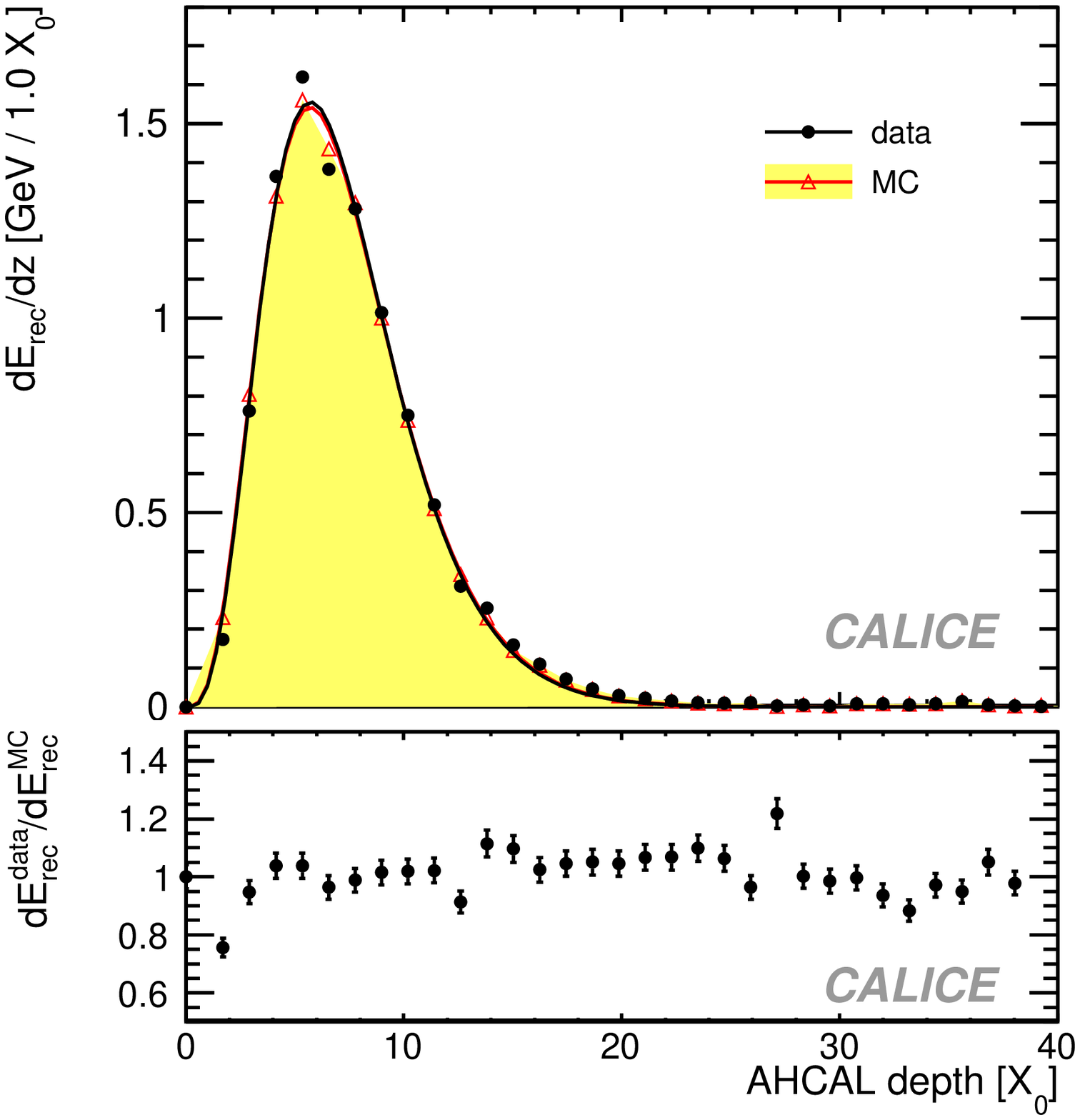}
    \includegraphics[width=0.49\textwidth]{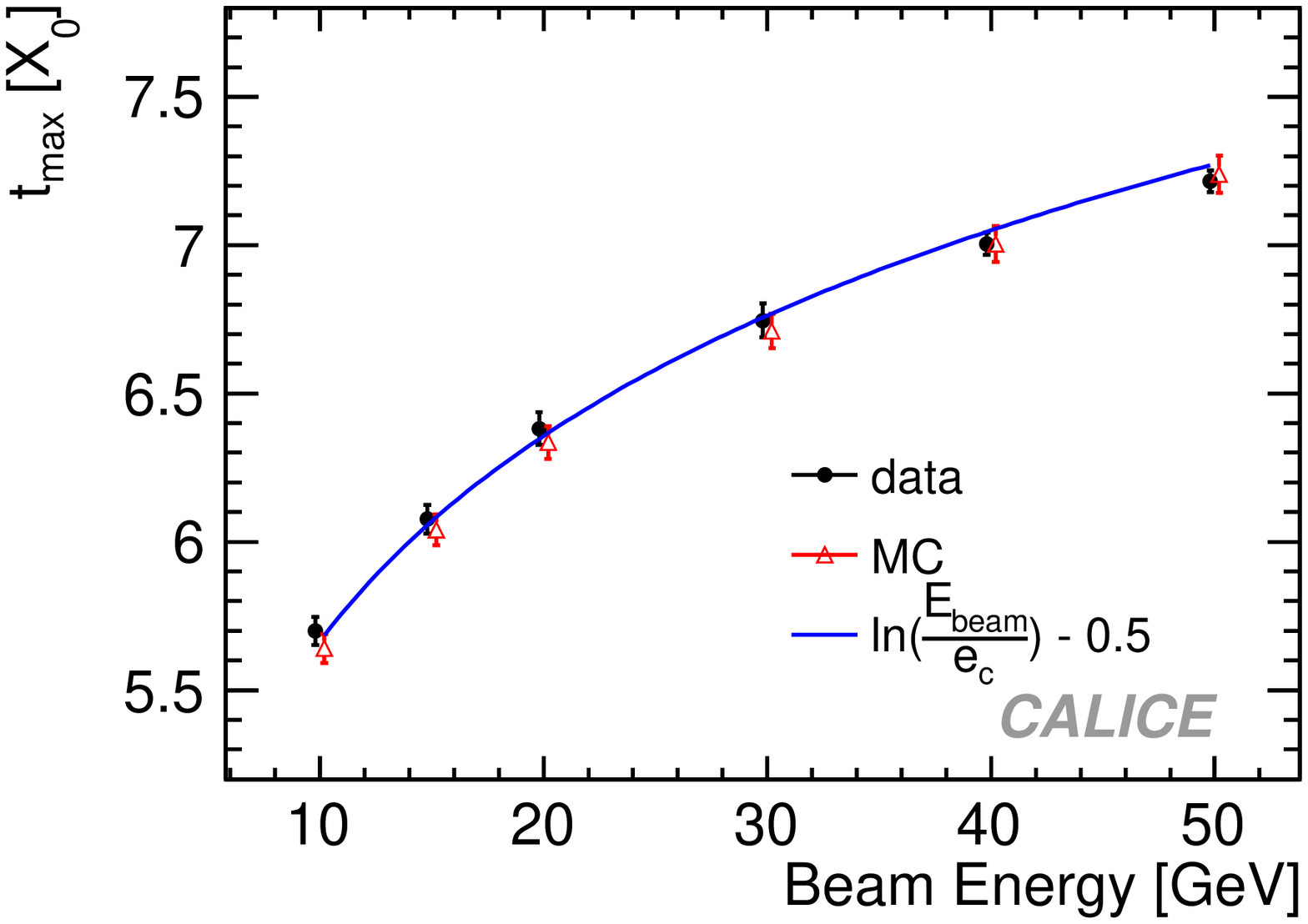}
 \end{center}
 \caption[]{{Longitudinal profile of a 10\,GeV positron shower in units of $X_0$ (left) and 
     scaling of the shower maximum as a function of the incident energy (right).
     The reconstructed energy (left plot) is shown for data (solid points), simulation (shaded area) and a fit to the data using Eq.~4.2 (line). The bottom insert shows the data/Monte Carlo comparison. The shower maximum (right plot) is shown for data (dots), simulation (open triangles) and the theory expectation given in Eq.~4.3 (solid line).}}
  \label{fig:profile}
\end{figure}

The mean longitudinal profile of a 10\,GeV positron shower is shown in the left plot of 
Figure~\ref{fig:profile}. Due to the high longitudinal segmentation of
the AHCAL, the shower rise, maximum and decay are clearly visible.
Data and simulation are in qualitatively good agreement. To quantify this
agreement, the profiles at each recorded beam energy 
are fitted with Eq.~\ref{eq:profile} and the maximum shower depth calculated as 
$ t_{\rm{max}} = \omega/b$.
The development of the shower maximum as a function 
of the beam energy is shown in the right plot of Figure~\ref{fig:profile}. The
error bars show the  uncertainty from the fits.
The extracted shower maxima of both data and simulation are in good
agreement with the theoretical behavior for a pure Fe calorimeter with a
critical energy, from~\cite{Leroy}, of $\epsilon_{\rm c} = 21.04$\,MeV, given in
Eq.~\ref{eq:t_max}.

\begin{figure}[!t]
  \begin{center}
    \includegraphics[width=0.55\textwidth]{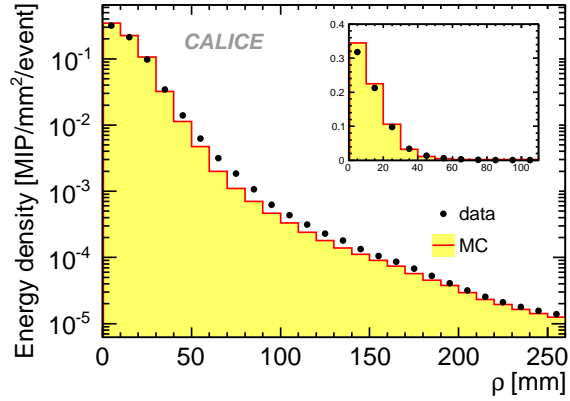}
 \end{center}
 \caption[]{{Transverse profile of a 15 GeV positron shower. The energy density is shown in 10 mm wide concentric rings centered around the shower axis.}}
  \label{fig:radius}
\end{figure}

The transverse shower profile of a 15 GeV positron shower is shown in Figure~\ref{fig:radius} together with a simulation. The radius, $\rho$, is calculated with respect to the track of the incoming particle extrapolated from the tracking system to the AHCAL front face. Therefore, the radius is defined as $\rho^2_i = (x_i - x_{\rm track})^2+  (y_i - y_{\rm track})^2$, where ($x_i,y_i$) are the coordinates of the calorimeter cell with signal above threshold. The energy deposited in a calorimeter cell is normally assigned to the center of the cell. For the radial profile studies it is redistributed uniformly in bins of 1~mm$^2$ before being assigned to one annular bin of inner radius $\rho$. In this way the energy deposited in one calorimeter cell can be shared between two adjacent annular bins. Proper normalization accounts for the fraction of the calorimeter cell area covered by each annular bin.
\begin{figure}[!b]
  \begin{center}
    \includegraphics[width=0.49\textwidth]{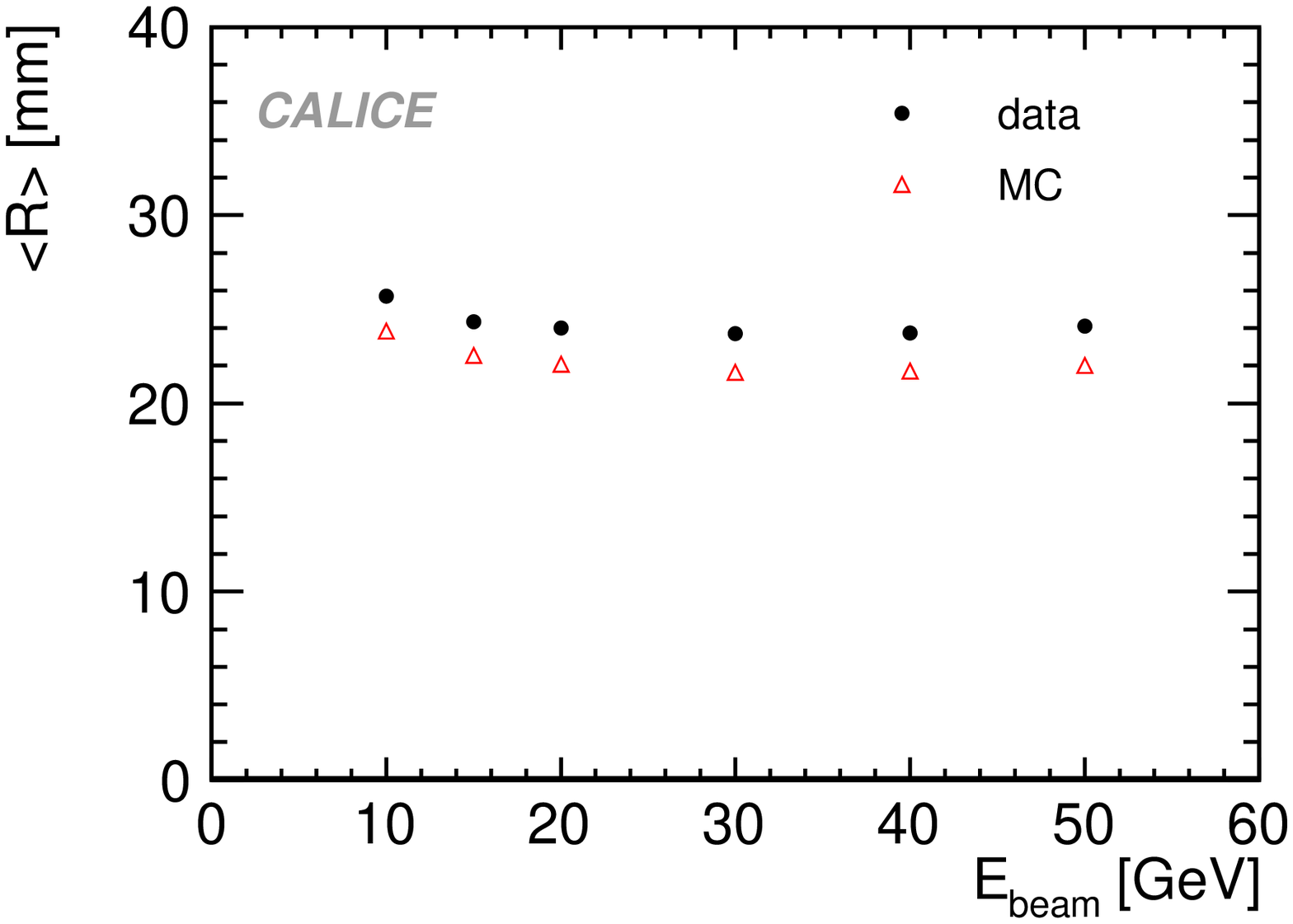}
    \includegraphics[width=0.49\textwidth]{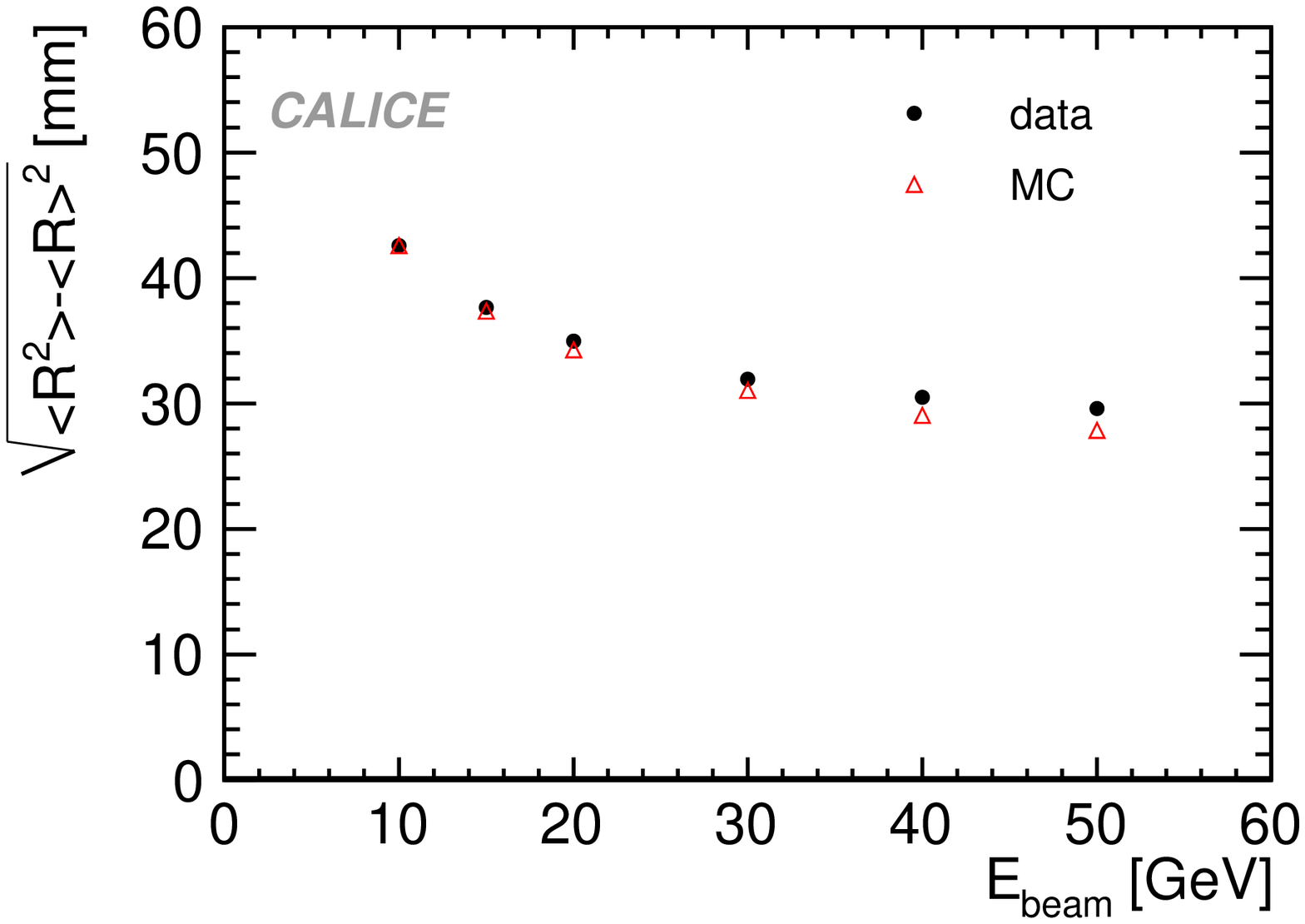}
 \end{center}
 \caption[]{{Mean (left) and RMS (right) of the transverse shower distribution as a function of beam energy. Dots are from data and open triangles are from simulation.}}
  \label{fig:R_RMS}
\end{figure}
The data indicate a broader shower than expected from simulation. 
The calculated mean shower radius ($\langle R \rangle = \frac{\Sigma E_i \rho_i}{\Sigma E_i}$) for 15 GeV showers in data is about 9\,\% larger than the simulated one. 

The energy dependence of $\langle R \rangle$ is shown on the right plot of Figure~\ref{fig:R_RMS} (left). The difference is almost energy independent. For completeness also the comparison of the RMS ($\sqrt{\langle R^2 \rangle - \langle R \rangle^2}$) of the shower radius distribution is shown in Figure~\ref{fig:R_RMS} (right).  An energy dependent disagreement of data and Monte Carlo is observed for this variable which increases to a maximum of 7\,\% for 50 GeV.
Several studies have been performed to find the cause of this effect including noisy and inactive cells, different beam shape, influence of the light cross-talk between tiles, misalignment between calorimeter and tracking system and of calorimeter layers. 
The broader shower in data is still not understood and further studies of asymmetric light collection on the tile, the influence of varying dead space between tiles due to varying thickness of reflector coating, etc., will follow to investigate the discrepancy.
For the purpose of the validation of the calibration procedure the current level of agreement is acceptable, though this mismatch will have to be taken into account when comparing hadronic shower shapes.
Furthermore, hadronic showers have a much smaller energy density than electromagnetic showers; therefore, any local effects, (i.e. the impact of dead areas or misalignment between layers), are strongly amplified in electromagnetic showers, while the influence is expected to be much less pronounced in hadronic showers.


\section{Uniformity studies}\label{sec:uniformity}
\subsection{Uniformity of the calorimeter response}\label{sec:calo_uniformity}
\begin{figure}[!t]
  \begin{center}
    \includegraphics[width=0.4\textwidth]{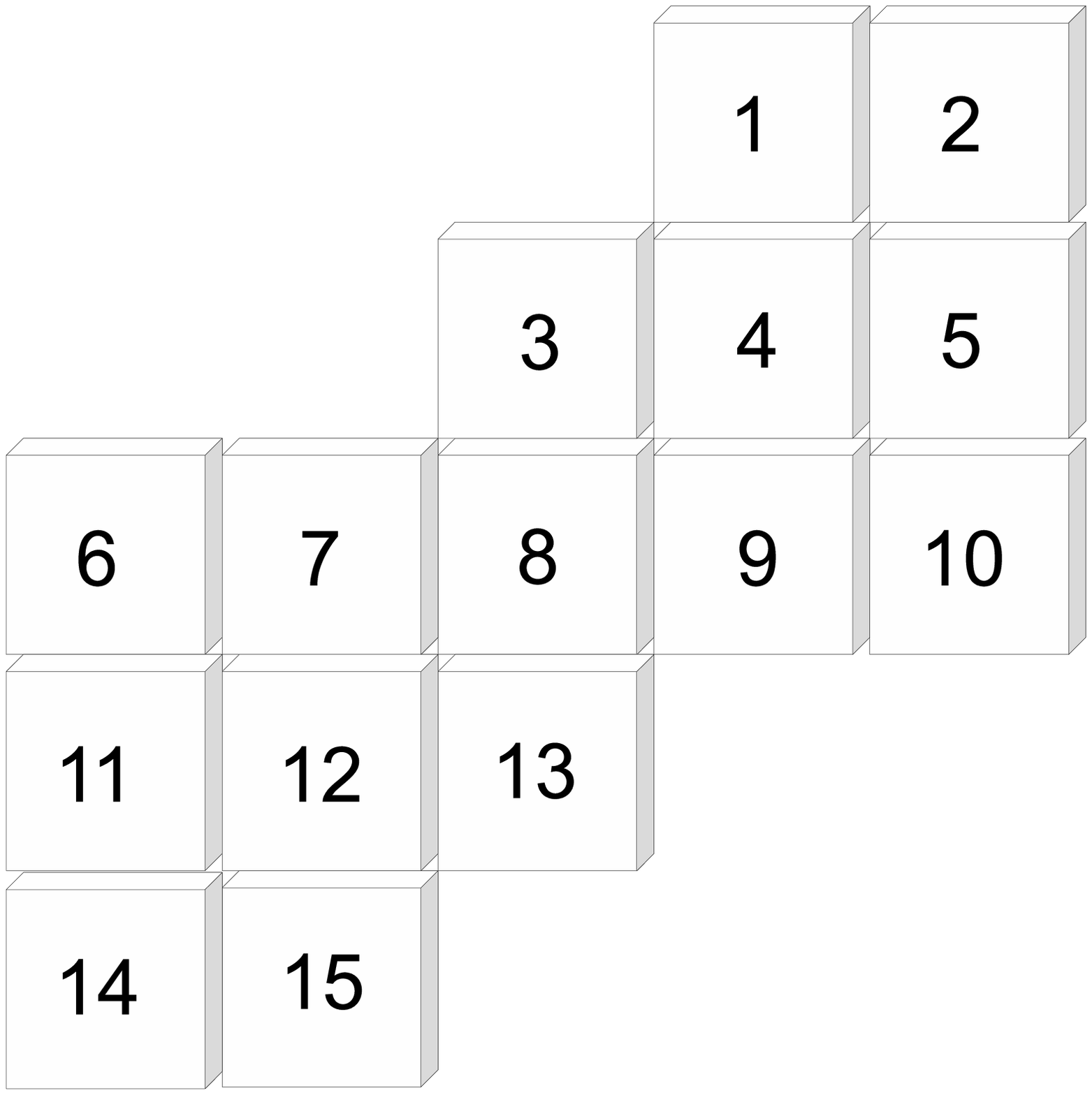}
    \includegraphics[width=0.55\textwidth]{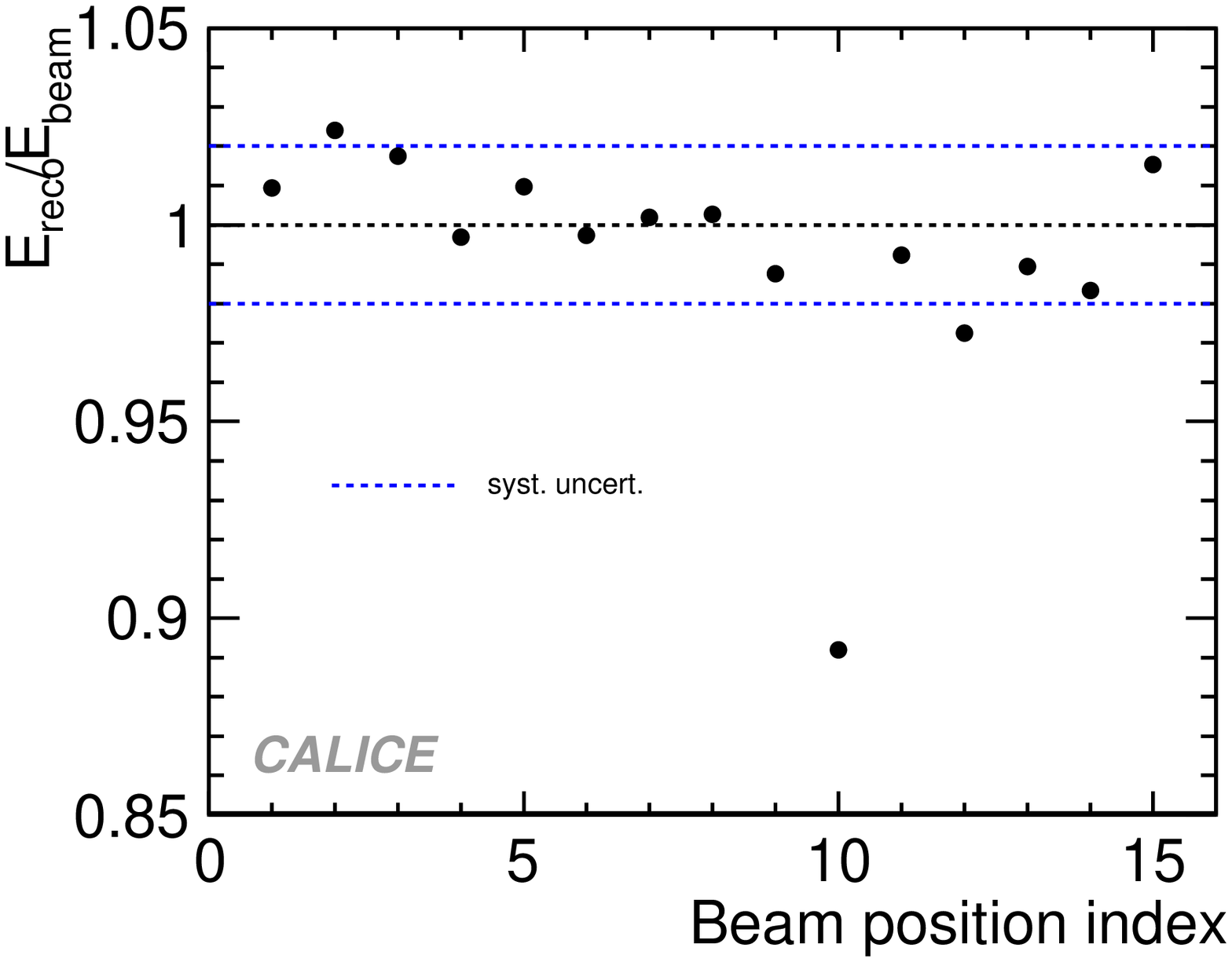}
 \end{center}
 \caption[]{{ Schematic view of tile positions in an AHCAL scintillator plane used for the uniformity test (left) and uniformity of the calorimeter response for various positions of incident beam with respect to the detector (right). Tile position eight is approximately in the center of each calorimeter layer. The dashed lines show the systematic uncertainties. Statistical uncertainties are negligible.}}
  \label{fig:uniformity}
\end{figure}

The uniformity of the AHCAL response is explored by shifting the AHCAL
to different positions with respect to the beam axis, at normal
angle of incidence. This procedure is visualized in the left part of
Figure~\ref{fig:uniformity}. Each square in the sketch represents one
scintillating tile of $3\times3$\,cm$^2$ and beam events with a track pointing to a 1$\times$1 cm$^2$ region centered on each tile in turn were selected. The uniformity of
the calorimeter response at the 15 different positions has been tested.
For this study 10\,GeV positron runs are analyzed, where the 
movable stage was used to displace the calorimeter in the $x$-$y$ position
with respect to the beam-line ($z$-axis). 

As shown in the right plot of Figure~\ref{fig:uniformity}, when excluding position~10, 
the uniformity of the calorimeter response is better than 2.1\,\%. 
The 10\,\% deviation between reconstructed and beam energy at position~10 is due to an inactive cell at the shower maximum,
which is not corrected in the calibration.

\subsection{Angular dependence of the calorimeter response}\label{sec:angle_uniformity}

\begin{figure}[!t]
  \begin{center} 
    \includegraphics[width=0.44\textwidth]{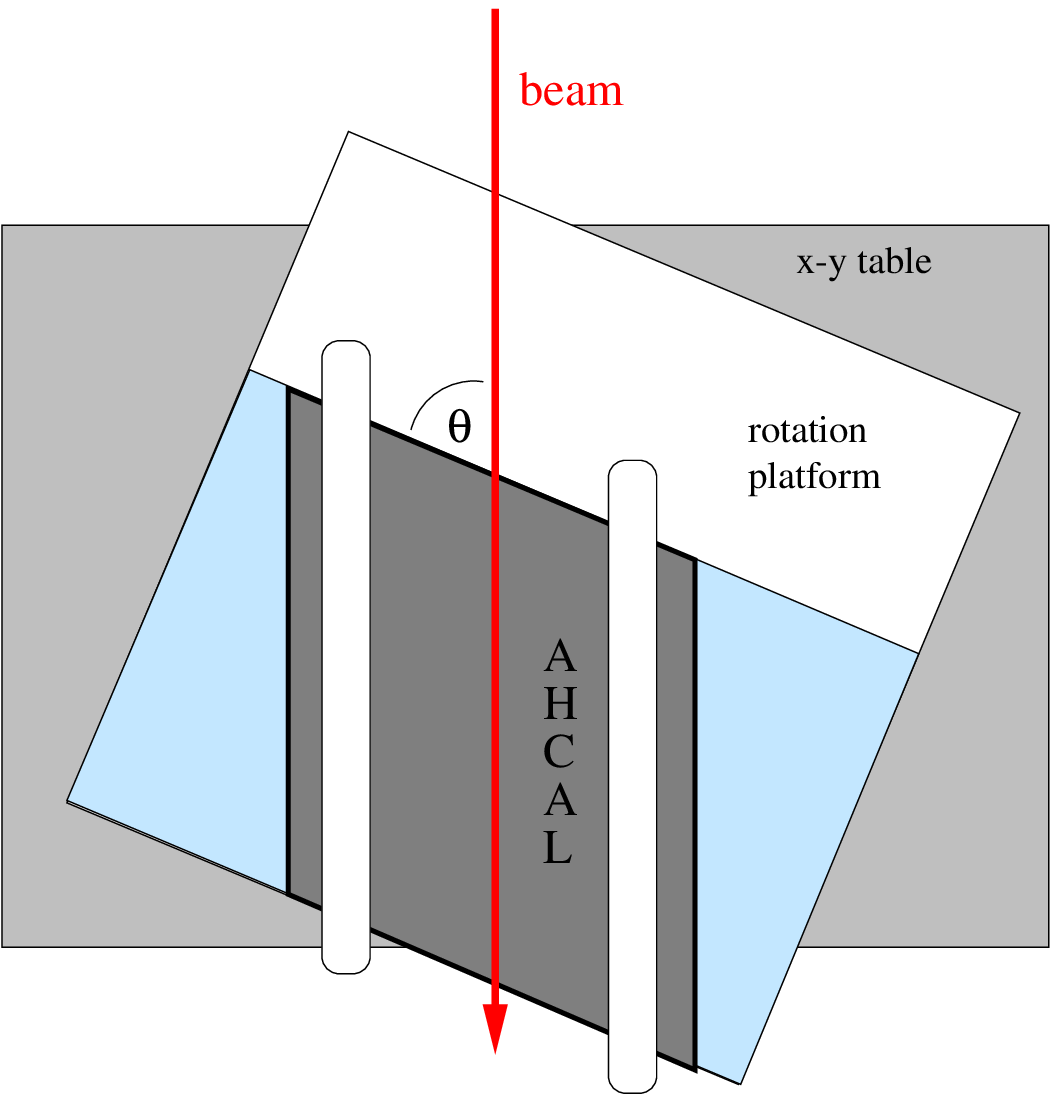}
    \includegraphics[width=0.55\textwidth]{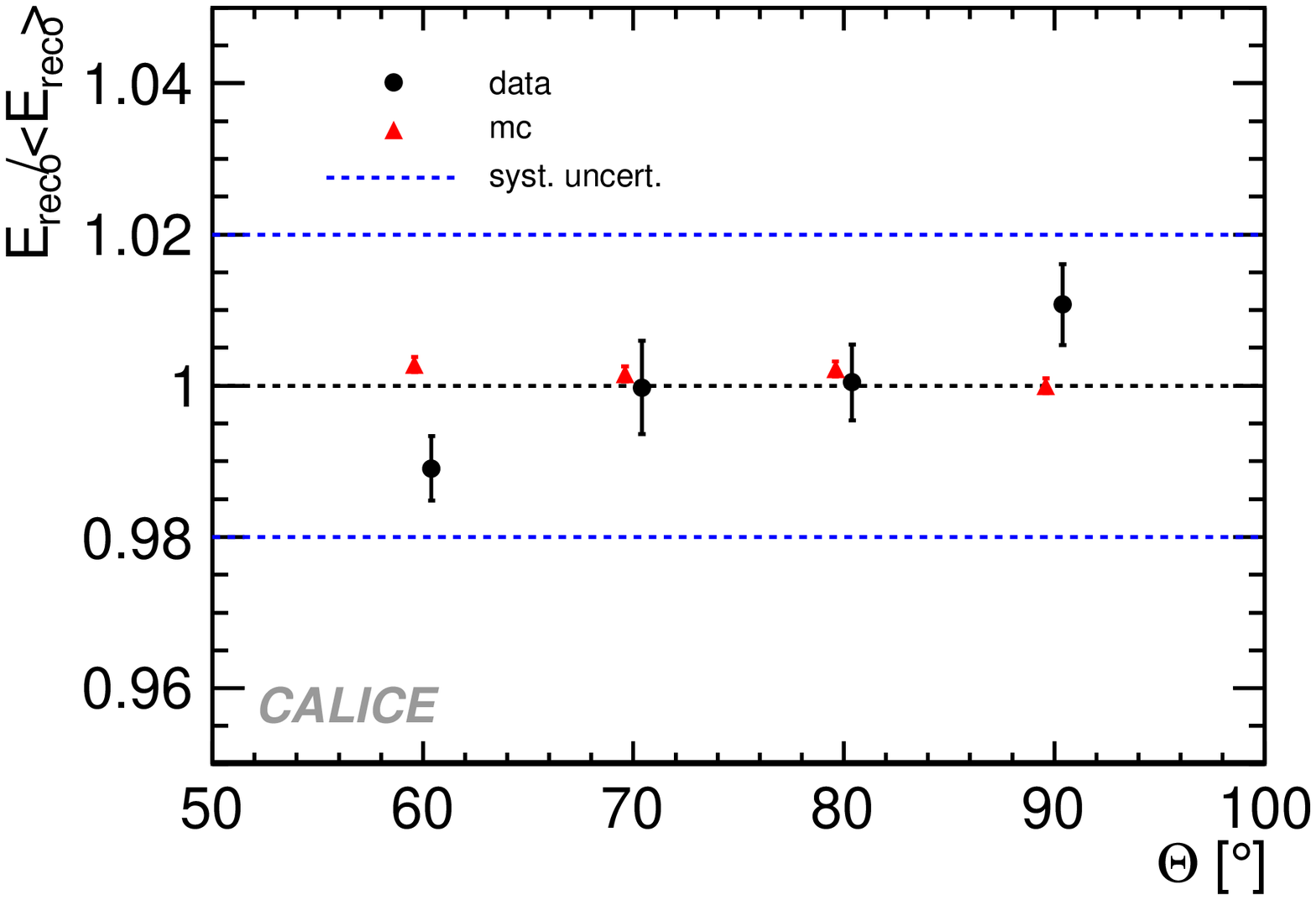}
 \end{center}
 \caption[]{{  Schematic view of the AHCAL rotated with respect to the beam (left) and reconstructed energy of 10 GeV positrons normalized to the average versus angle of incidence (right). To improve legibility, the data (solid points) and the simulation (red triangles) are slightly shifted in opposite directions on the abscissa. The systematic uncertainty is shown by dash-dotted lines.  
 Additionally, the spread of all measurements performed at one inclination angle are shown 
     as an error for each point.}}
  \label{fig:angular_dependence}
\end{figure}

The movable stage carrying the AHCAL was used to collect
positron data  at incident angles of $90^\circ$,  
$80^\circ$, $70^\circ$ and $60^\circ$. 
The rotation and staggering of the AHCAL are sketched in the left plot of
Figure~\ref{fig:angular_dependence}, where the beam is entering from
the top. In the rotated configuration, the modules
were staggered to ensure the highly granular core of 3$\times$3 cm$^2$ 
was aligned with, and hence sampled, the shower core.

For each angle of incidence $\theta$ several 10\,GeV positron runs with different impact points on the calorimeter front surface were taken. The average of all the runs is used to define $E_{reco}$ at one given $\theta$, while $\langle E_{reco} \rangle$ is the mean reconstructed energy at any angle of incidence. 
The normalized reconstructed energy is plotted in Figure~\ref{fig:angular_dependence} as a function of $\theta$.
The spread (RMS) between the various analyzed runs per inclination angle is used as the systematic uncertainty. 
This spread is smaller than the calibration systematic uncertainty in the calorimeter, shown in the plot as an error band around the ratio of one. Showers at various inclination angles only partially share the same calorimeter cells, therefore the full systematic uncertainty from calibration is an overestimate of the real error, but the spread between  measurements performed at one inclination angle is an underestimate. 
Taking this into account, the increase in the reconstructed energy of data between $60^\circ$ and $90^\circ$ is not significant. A more precise analysis would require more data at different angles which are not available at present.

\subsection{Influence of cell structure}\label{sec:cell_uniformity}
\begin{figure}[!t]
  \begin{center}
    \includegraphics[width=0.41\textwidth]{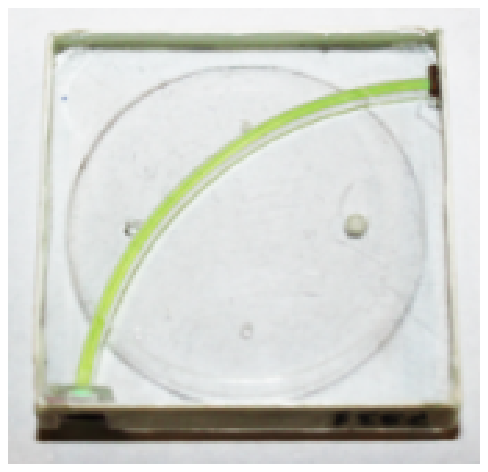}
    \includegraphics[width=0.58\textwidth]{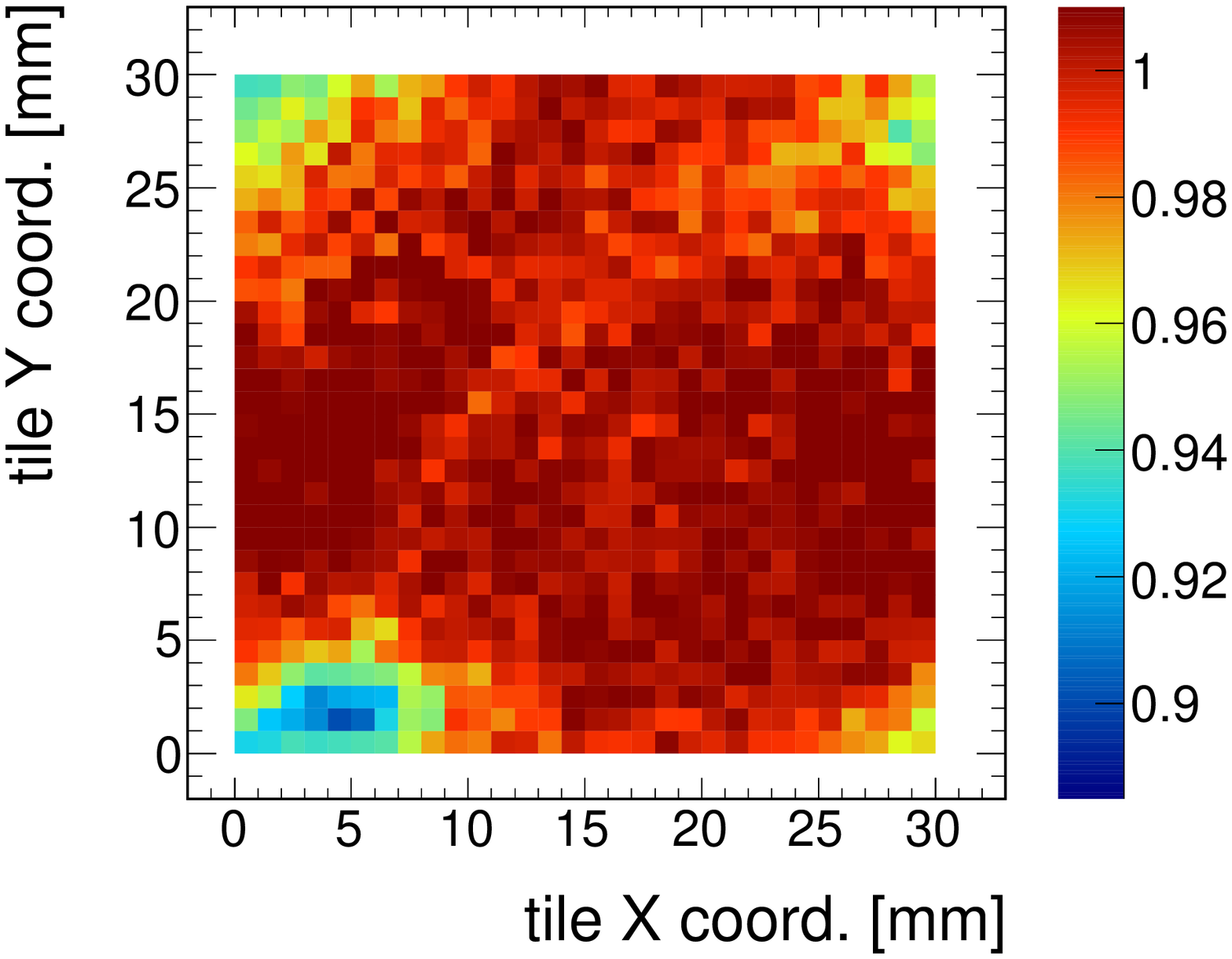}
 \end{center}
 \caption[]{{Picture
     of the scintillating tile (left).
     Effect of the AHCAL scintillator tile structure on the
     energy measurement (summed over the entire calorimeter) for 10 GeV electromagnetic showers (right).}}
  \label{fig:homogeneity}
\end{figure}

The scintillating tiles used in the AHCAL have a WLS fiber embedded in
a groove, a SiPM inserted into a small groove on one end of this fiber,
and a mirror in a groove on the other end. This structure is visible
in the picture of a $3\times3$\,cm$^2$ tile in
Figure~\ref{fig:homogeneity}, where the SiPM is located in the lower
left-corner, the mirror is placed in the diagonal corner, and the WLS
fiber is embedded in a quarter circle. 

Since electromagnetic showers have a short transverse extension, the impact of the
cell structure slightly reduces the energy resolution of the
calorimeter. To study this effect, we take advantage of the delay wire
chambers that were present in the beam line. They are used to
reconstruct the track of the incoming particle. This track is then
extrapolated to the front face of the AHCAL. 
The shower energy (energy summed over the entire calorimeter) 
for 10 GeV positrons with this impact position, normalized to the 
shower energy averaged over all impact positions, 
is plotted in Figure~\ref{fig:homogeneity}.

As shown in the figure, the measured energy drops slightly over the area of the WLS fiber.
A particle with a trajectory intersecting the SiPM (in the lower left corner of the plot) or the reflecting mirror at the end of the WLS fiber (in the upper right corner of the plot) shows a significant loss of response  with respect to the tile average by about 8\,\% and 4\,\% respectivley. At the position of the WLS fiber the tile response is about 2\,\% lower than average. The drop at the other two corners of the tile in this study reflects the energy loss associated with the SiPMs located in the neighboring tiles as the observable used is the energy summed over the entire calorimeter.
Measurements of single tile uniformity using a collimated source have been performed and are reported in~\cite{uniformity1,uniformity}. These measurements confirm a lower response of electromagnetic showers hitting the SiPMs or the reflecting mirrors. Though this large degradation (8\,\% at the locations of the SiPMs) is quite unrealistic in a collider detector, where the particles 
are always traversing the calorimeter under an angle. In this case 
the tile response non-uniformity averages out with no influence on the energy resolution.
Furthermore, electromagnetic showers have a short lateral extension. 
For pion showers, which are much wider, the effect has not been observed in data. 
The effects of gaps between the calorimeter tiles, as well as the non-uniform response of the tiles, in view of the impact on the energy resolution, have been studied using Monte Carlo events. The results are reported in~\cite{uniformity2} and show that these type of effects do not have a significant influence on the measurement of hadron showers.


\section{Conclusions}\label{sec:conclusion}
The response of the CALICE analog
hadron calorimeter to positrons was measured for energies between 10 and 50\,GeV,
using data recorded at CERN in summer 2007. The calorimeter
response is linear to better than 3\,\%. 
A better SiPM saturation correction would improve the linearity, and for future developments a larger dynamic range is desirable.
This study is ongoing, but the effect on pion energy 
reconstruction will be negligible due to the much smaller energy per 
hit in a hadronic shower compared to an electromagnetic shower.
The energy resolution for positrons is found to have a stochastic term of  
$(21.9 \pm 1.4)$\,\%$\sqrt{E\,[\rm{GeV}]}$, and a constant term of about 1\,\%. 
Good agreement between data and simulation validates the 
simulation of the various detector characteristics.
For comparison, Ref.~\cite{ATLAS} reports for the ATLAS tile calorimeter an energy resolution of 28\,\%$ \sqrt{E\,[\rm{GeV}]}$ stochastic and  2.8\,\% constant term for electrons at 20 deg from normal incidence. This is also and hadron sampling calorimeter alternating steel and scintillator tiles, but with a much coarser granularity than the AHCAL and a different a readout via standard photomultiplier tubes.
The same readout technology as in the AHCAL is also implemented in a scintillator-Tungsten electromagnetic calorimeter, ScECAL build within the CALICE collaboration~\cite{ScECAL}. For this calorimeter the energy resolution to electrons is of  $(15.15 \pm 0.03)$\,\%$\sqrt{E\,[\rm{GeV}]}$ stochastic and $(1.44 \pm 0.02)$\,\% constant term. This analysis provided confidence that the detector performance and simulation
are sufficiently understood to pursue the investigation of hadronic showers.

Systematic studies are performed to investigate the quality of the calibration in as 
 many calorimeter cells as possible. 
The uniformity of the calorimeter response to electromagnetic showers is studied with beams at different impact points and different incident angles.
The results are consistent with no angular and spatial dependence within the quoted systematic uncertainty 
on the calibration procedure.

The high  segmentation of the AHCAL is well-suited for studying the longitudinal
 shower development with high accuracy and for determining the shower maximum. 
The point of maximum energy deposition along the shower propagation axis is 
located between 5.5\,$X_0$ and 7\,$X_0$ for the range of particle energies used,
consistent with simulation and theoretical prediction.

The transverse shower spread is more difficult to measure because it is strongly affected by 
uncertainties in the beam profile, in the variation of light cross-talk between tiles, 
and in the misalignment of calorimeter layers. Currently, the data indicate a broader 
shower than expected from simulation. However, the level of agreement is acceptable for the  
validation of the calibration procedure if one considers that the effect on hadronic showers
will be less important due to the lower energy density of hadronic showers.


\acknowledgments

We would like to thank the technicians and the engineers who
contributed to the design and construction of the prototypes. We also
gratefully acknowledge the DESY and CERN managements for their support and
hospitality, and their accelerator staff for the reliable and efficient
beam operation. 
This work was supported by the 
Bundesministerium f\"{u}r Bildung und Forschung (BMBF), grant no. 05HS6VH1, Germany;
by the DFG cluster of excellence `Origin and Structure of the Universe' of Germany ; 
by the Helmholtz-Nachwuchsgruppen grant VH-NG-206;
by the Alexander von Humboldt Foundation (Research Award IV, RUS1066839 GSA);
by joint Helmholtz Foundation and RFBR grant HRJRG-002, SC Rosatom;
by Russian Grants SS-3270.2010.2 and RFBR08-02-12100-OFI and by Russian
Ministry for Education and Science; by MICINN and CPAN, Spain;
by CRI(MST) of MOST/KOSEF in Korea;
by the US Department of Energy and the US National Science
Foundation;
by the Ministry of Education, Youth and Sports of the Czech Republic
under the projects AV0 Z3407391, AV0 Z10100502, LC527  and LA09042  and by the
Grant Agency of the Czech Republic under the project 202/05/0653;  
and by the Science and Technology Facilities Council, UK.


\end{document}